\documentclass[10pt,twocolumn,pra,aps,showkeys,showpacs]{revtex4-1}
\usepackage[latin9]{inputenc}
\usepackage[svgnames]{xcolor}
\usepackage{pdfcolmk}
\usepackage{mathrsfs}
\usepackage{amsmath}
\usepackage{amssymb}
\usepackage{graphicx}
\usepackage{MnSymbol}
\usepackage{soul,xcolor}

\usepackage[T1]{fontenc}
\usepackage[unicode=true,
 bookmarks=false,
 breaklinks=false,pdfborder={0 0 1},colorlinks=true]
 {hyperref}



\begin{document}
\setstcolor{red}

\title{Floquet analysis of a quantum system with modulated periodic driving}

\author{Viktor Novi\v{c}enko}

\affiliation{Institute of Theoretical Physics and Astronomy, Vilnius University,
Saul\.{e}tekio Ave.~3, LT-10222 Vilnius, Lithuania}

\author{Egidijus Anisimovas}

\affiliation{Institute of Theoretical Physics and Astronomy, Vilnius University,
Saul\.{e}tekio Ave.~3, LT-10222 Vilnius, Lithuania}

\author{Gediminas Juzeli\={u}nas}

\affiliation{Institute of Theoretical Physics and Astronomy, Vilnius University,
Saul\.{e}tekio Ave.~3, LT-10222 Vilnius, Lithuania}

\begin{abstract}
We consider a 
quantum system periodically driven with a strength which varies slowly
on the scale of the driving period. The analysis is based on a general
formulation of the Floquet theory relying on the extended Hilbert
space. It is shown that the dynamics of the system can be described
in terms of a slowly varying effective Floquet Hamiltonian that captures
the long-term evolution, as well as rapidly oscillating micromotion
operators. We obtain a systematic high-frequency expansion of all
these operators. Generalizing the previous studies, the expanded
effective Hamiltonian is now time-dependent and contains extra terms
appearing due to changes in the periodic driving. The same applies
to the micromotion operators which exhibit a slow temporal dependence
in addition to the rapid oscillations. As an illustration, we consider
a quantum-mechanical spin in an oscillating magnetic field with a
slowly changing direction. The effective evolution of the spin is
then associated with non-Abelian geometric phases reflecting
the geometry of the extended Floquet space. The developed formalism
is general and also applies to other periodically driven systems,
such as shaken optical lattices with a time-dependent shaking strength,
a situation relevant to the cold atom experiments. 
\end{abstract}
\maketitle

\section{Introduction\label{sec:Introduction}}

In recent years there has been a growing interest in periodically
driven quantum systems. The current surge of activities stems, to
a considerable extent, from a possibility to control and alter the
topological~\cite{Oka2009,Kitagawa2010,Galitski2011NP,ho12transport,hauke13,Rudner2013,Ketterle:2013,Aidelsburger:2013,Jotzu2014,Galitski13PRB,grushin14,Aidelsburger14NP,baur14,Zhai2014-Floquet,zhang-zhou14,Mei14,Anisimovas15PRB,zheng15fflo,verdeny15,przys15,dutta16,xiong16chern,bomantara16,plekhanov16,Flaschner16,flaeschner16b}
and many-body~\cite{sorensen05,eckardt05,zenesini09,Eckardt2010,neupert11,regnault11,struck11,wu12,Lewenstein2012,Struck:2013,bergholtz13,parameswaran13,Greschner14,raciunas16,Nagerl2016,Eckardt16-Review}
properties of the systems by periodically driving them~\cite{Arimondo2012,Windpassinger2013RPP,Eckardt16-Review,Goldman2014,Goldman2014RPP,Bukov2015,goldman15resonant,Eckardt2015,itin15,Mikami16PRB,holthaus16tutorial,hemmerich10,kolovsky11,creffield13comment,creffield14,Hauke:2012,iadecola15,jimenez15,Heinisch16}.
This extends to a broad range of condensed matter~\cite{Oka2009,Galitski2011NP,Kitagawa2011,Galitski13PRB,tong13majorana,grushin14,usaj14,quelle15},
photonic~\cite{Haldane:2008cc,Rechtsman:2013fe} and ultracold atom~\cite{Eckardt16-Review,Goldman2014RPP,Windpassinger2013RPP,Dalibard2011,Aidelsburger:2011,Aidelsburger:2013,atala14,zenesini09,Aidelsburger14NP,Arimondo2012,struck11,struck12,Hauke:2012,hauke13,Struck:2013,Jotzu2014,Ketterle:2013,Kennedy15,Flaschner16,Budich16,Nagerl2016,Galitski2013,Zhai2012JMPB,Zhai2014-review,Anderson2013,Xu2013,jimenez15,nascimbene15,perez-piskunow15,Luo16Sci_Rep}
systems. An important situation arises when the driving frequency
exceeds other characteristic frequencies of the system. In that case,
one can construct a high-frequency expansion of an effective time-independent
Hamiltonian of the system in the inverse powers of the driving frequency~\cite{Rahav03,Guerin2003,Arimondo2012,Goldman2014,goldman15resonant,Eckardt2015,Bukov2015,itin15,Mikami16PRB,Eckardt16-Review,Heinisch16,holthaus16tutorial}.
In addition to the long-term dynamics represented by such an effective
(Floquet) Hamiltonian there is also a fast modulation on the scale
of a single driving period described by the micromotion operators.

In many cases, the periodic driving is changing within an experiment.
Here, we provide a general analysis of a behavior of such a quantum
system subjected to a high-frequency perturbation which additionally
changes in time. The analysis is based on a general formulation of
the Floquet theory using an extended space approach~\cite{sambe73,Howland1974,Breuer1989,Breuer1989a,Grifoni98,Drese1999,Eckardt08,Guerin2003,Heinisch16}.
In addition to a fast periodic modulation, we allow the Hamiltonian
to have an extra (slow) time dependence. We show that the dynamics
of the system can then be factorized into the following contributions:
(i) a long-term evolution is determined by a slowly varying effective
Floquet Hamiltonian; (ii) rapid oscillations are described by micromotion
operators which are additionally slowly changing in time. This factorization
represents an extension of the Floquet approach to Hamiltonians which
are not entirely time-periodic. Note that an exponential form of the
slowly varying effective evolution operator now involves time ordering
if the effective Hamiltonian does not commute with itself at different
times.

We obtain a high-frequency expansion of the effective Hamiltonian
and micromotion operators. Generalizing 
the previous studies~\cite{Eckardt2015,Goldman2014,Bukov2015,Mikami16PRB,Eckardt16-Review},
the expanded effective Hamiltonian is now time dependent and contains
extra terms due to the changes in the periodic driving. The same applies
to the micromotion operators which exhibit a slow temporal dependence
in addition to the rapid oscillations. 

The theory is illustrated by considering a spin in an oscillating
magnetic field with a slowly changing direction. In that case, the
effective evolution of the spin is associated with non-Abelian (non-commuting) geometric phases if the oscillating magnetic field is not restricted to a single
plane. The formalism can be applied to describe other periodically
driven systems, such as shaken optical lattices with a time-dependent
shaking strength, which are relevant to the cold atom experiments~\cite{Eckardt2010,Aidelsburger:2011,Arimondo2012,Hauke:2012,Windpassinger2013RPP,Struck:2013,Aidelsburger:2013,Jotzu2014,Aidelsburger14NP,Kennedy15,Flaschner16}.

The paper is organized as follows. In the following Sec.~\ref{subsec:Formulation} we formulate the problem and review the basic elements of the Floquet formalism which underpin the subsequent generalization of the approach to the case of slowly modulated driving. In Sec.~\ref{sec:Evolution and high frequency expansion} we consider the temporal evolution   of the periodically driven system taking into account of the slow modulation of the driving, as well as present the high-frequency expansion of the effective Hamiltonian and micromotion operators describing such an evolution. In Sec.~\ref{sec:Oscillating spin} the general formalism is applied to the spin in an oscillating
magnetic field with a slowly changing direction.  The concluding Sec.~\ref{Conclusions}
summarizes the findings. Details of some calculations and other auxiliary material are presented in four appendixes. In particular, Appendix \ref{sec:1d-lattice} analyzes the Floquet effective Hamiltonian for a one-dimensional shaken optical lattice with a slowly changing amplitude of driving. 

\section{Problem formulation and background material\label{subsec:Formulation}}

\subsection{Hamiltonian and equations of motion}

Let us consider the time evolution of a quantum system described by
a Hamiltonian $H\left(\omega t+\theta,t\right)$ which is $2\pi$-periodic
with respect to the first argument

\begin{equation}
H\left(\omega t+\theta,t\right)=H\left(\omega t+\theta+2\pi,t\right)\,,\label{eq:H-periodic}
\end{equation}
where an angle $\theta$ defines an initial phase of the Hamiltonian.
A possibility to have an additional temporal dependence (not necessarily
periodic) is represented by the second argument $t$. We will address the situation where the first argument in $H\left(\omega t+\theta,t\right)$
describes fast temporal oscillations, whereas the second argument
plays the role of a slowly varying envelope.

The Hamiltonian $H\left(\omega t+\theta,t\right)$ can be expanded
in a Fourier series with respect to the first argument 
\begin{equation}
H\left(\theta^{\prime},t\right)=\sum_{l=-\infty}^{\infty}H^{\left(l\right)}\left(t\right)e^{il\theta^{\prime}}\,,\quad\theta^{\prime}=\omega t+\theta\label{eq:H-periodic-expansion}
\end{equation}
where the expansion components $H^{\left(l\right)}\left(t\right)$
are generally time-dependent operators. In this way, the second argument
in $H\left(\theta^{\prime},t\right)$ represents the temporal modulation
of the amplitudes $H^{\left(l\right)}\left(t\right)$ of the harmonics
$e^{il\theta^{\prime}}$. Since the Hamiltonian is Hermitian, the
negative frequency Fourier components are Hermitian conjugate to the
positive frequency ones: $H^{\left(l\right)\dagger}\left(t\right)=H^{\left(-l\right)}\left(t\right)$.

The quantum state of the system is described by a state-vector $\left|\phi_{\theta}\left(t\right)\right\rangle $
obeying a time-dependent Schr\"{o}dinger equation (TDSE): 
\begin{equation}
i\hbar\partial_{t}\left|\phi_{\theta}\left(t\right)\right\rangle =H\left(\omega t+\theta,t\right)\left|\phi_{\theta}\left(t\right)\right\rangle .\label{eq:Shroedinger-initial}
\end{equation}
The subscript $\theta$ in the state vector $\left|\phi_{\theta}\left(t\right)\right\rangle $
appears, because the dynamics of the state vector $\left|\phi_{\theta}\left(t\right)\right\rangle $
is governed by the Hamiltonian $H\left(\omega t+\theta,t\right)$
which parametrically depends on the phase $\theta$. Therefore, the
state vector $\left|\phi_{\theta}\left(t\right)\right\rangle $ evolves
differently for different phases $\theta$ entering the Hamiltonian
$H\left(\omega t+\theta,t\right)$ even though $\left|\phi_{\theta}\left(t\right)\right\rangle $
is $\theta$-independent at the initial time $t_{0}$ 
\footnote{Subsequently in Eq.~(\ref{eq:initial-condition-theta-independent})
we shall adopt such a $\theta$-independent initial condition.}.

Since the Hamiltonian $H\left(\omega t+\theta,t\right)$ is $2\pi$-periodic
with respect to its phase $\theta$, one can choose the state vector
$\left|\phi_{\theta}\left(t\right)\right\rangle $ to have the same
$\theta$ periodicity: 
\begin{equation}
\left|\phi_{\theta+2\pi}\left(t\right)\right\rangle \equiv\left|\phi_{\theta}\left(t\right)\right\rangle .\label{eq:state-vector-periodic-cond}
\end{equation}
Such a state vector can be expanded in terms of a Fourier series 
\begin{equation}
\left|\phi_{\theta}\left(t\right)\right\rangle =\sum_{n=-\infty}^{\infty}\left|\phi^{(n)}\left(t\right)\right\rangle e^{in\theta},\label{eq:state-vector:_theta-expansion}
\end{equation}
where $\left|\phi^{\left(n\right)}\left(t\right)\right\rangle $ is
an $n$th harmonic (in the phase variable $\theta$) of the full
state vector $\left|\phi_{\theta}\left(t\right)\right\rangle $.

\subsection{Extension of the space}

The idea of extending the Hilbert space for periodical driven
systems goes back to a classical work by Sambe \cite{sambe73}. The
Floquet eigenstates are then obtained by solving a stationary Schr\"{o}dinger
equation governed by a time-independent Hamiltonian acting in the
expanded space. The role of the additional space is played by a temporal
variable, the periodic harmonics $e^{in\omega t}$ forming basis states
of the extra space. Subsequently, the approach has been extended to
incorporate temporal modulation of the periodic driving \cite{Howland1974,Breuer1989,Breuer1989a,Peskin1993,Grifoni98,Drese1999,Guerin2003,Fleischer2005,Eckardt08,Heinisch16}.
In particular, the analysis of periodically driven quantum systems
which contain slowly changing parameters has been initiated by Breuer
and Holthaus \cite{Breuer1989,Breuer1989a} using a two-time ($t,t^{\prime}$)
formalism. 

Here, we make use of another (yet equivalent) approach~\cite{Guerin2003}.
Specifically, we promote to a quantum variable the phase $\theta$
entering the Hamiltonian $H\left(\omega t+\theta,t\right)$, subsequently
eliminating the temporal dependence via a unitary transformation (\ref{eq:Unit})
acting in the extended Hilbert space. It is noteworthy that for a
particular value of $\theta$, the state vector $\left|\phi_{\theta}\left(t\right)\right\rangle $
is an element of the original (physical) Hilbert space $\mathscr{H}$,
and the Hamiltonian $H\left(\omega t+\theta,t\right)$ operates in
this space. On the other hand, for an arbitrary phase $\theta$ the
factors $e^{in\theta}$ featured in the state vector $\left|\phi_{\theta}\left(t\right)\right\rangle $
Eq.~(\ref{eq:state-vector:_theta-expansion}), can be treated as
an orthonormal set of basis vectors of an auxiliary Hilbert space
$\mathscr{T}$ comprising $\theta$-periodic functions in the interval
$\theta\in\left[0,2\pi\right)$. The inner product in $\mathscr{T}$
is defined as an integral $\left(2\pi\right)^{-1}\intop_{0}^{2\pi}e^{-im\theta}e^{in\theta}d\theta=\delta_{nm}$.
Thus, the state vector $\left|\phi_{\theta}\left(t\right)\right\rangle $
can be considered as an element of the extended Hilbert space $\mathscr{L}=\mathscr{H}\otimes\mathscr{T}$.
This approach corresponds to considering an evolution of an ensemble
of quantum systems governed by a set of Hamiltonians $H\left(\omega t+\theta,t\right)$
with various phases $\theta$. In order to distinguish between the state vectors in the spaces $\mathscr{H}$ and $\mathscr{L}$, we will use a convenient bra-ket notation $\langle \cdot|$ and $|\cdot \rangle$ for the physical space $\mathscr{H}$ and a double bra-ket notations $\llangle \cdot |$ and $|\cdot \rrangle$ for the extended space $\mathscr{L}$. Therefore the $\theta$-dependent physical state vector  $\left|\phi_{\theta}\left(t\right)\right\rangle $  will be labeled as  $\left|\phi_{\theta}\left(t\right)\right\rrangle $ if it is considered as an element of  $\mathscr{L}$. The operators acting in $\mathscr{H}$ are denoted without a hat like in Eq.~(\ref{eq:H-periodic}), whereas the operators acting in $\mathscr{L}$ will contain a hat over a symbol, such as  in Eq.~(\ref{eq:Unit}).

\subsection{Elimination of periodic temporal dependence in the extended space}

To eliminate the periodic temporal dependence of the Hamiltonian $H\left(\omega t+\theta,t\right)$
entering the TDSE~(\ref{eq:Shroedinger-initial}), let us apply a
unitary transformation in the extended space~\cite{Guerin2003}:\begin{subequations}
\label{eq:Unit} 
\begin{align}
\hat{U} & =\exp\left(\omega t\partial/\partial\theta\right),\label{eq:U}\\
\hat{U}^{-1} & =\hat{U}^{\dagger}=\exp\left(-\omega t\partial/\partial\theta\right).\label{eq:U^dagg}
\end{align}
\end{subequations}A hat over $\hat{U}$ signifies that it is an operator
acting in $\mathscr{L}$, as it contains a derivative $\partial/\partial\theta$. Due to the periodic boundary condition~(\ref{eq:state-vector-periodic-cond}) for the state vector
$\left|\phi_{\theta}\left(t\right)\right\rangle $ with respect to
$\theta$, the operator $-i\partial/\partial\theta$ is Hermitian
in the extended space. Consequently, the transformation $\hat{U}$
is unitary in $\mathscr{L}$.

The operator $\hat{U}=\hat{U}\left(\omega t\right)$ shifts the phase
variable: $\hat{U}^{\dagger}\theta\hat{U}=\theta-\omega t\,$, so
$\hat{U}^{\dagger}\hat{H}\left(\omega t+\theta,t\right)\hat{U}=\hat{H}\left(\theta,t\right)$
no longer has a fast periodic temporal dependence. The transformed
state vector 
\begin{equation}
\left|\psi_{\theta}\left(t\right)\right\rrangle =\hat{U}^{\dagger}\left|\phi_{\theta}\left(t\right)\right\rrangle \equiv\left|\phi_{\theta-\omega t}\left(t\right)\right\rrangle ,\label{eq:state-vector-transformed}
\end{equation}
obeys the TDSE 
\begin{equation}
i\hbar\partial_{t}\left|\psi_{\theta}\left(t\right)\right\rrangle =\hat{K}\left(\theta,t\right)\left|\psi_{\theta}\left(t\right)\right\rrangle ,\label{eq:Shroedinger-transformed}
\end{equation}
governed by a Hamiltonian $\hat{K}\left(\theta,t\right)=\hat{H}\left(\theta,t\right)-i\hbar\hat{U}^{\dagger}\partial_{t}\hat{U}$,
where an extra term is due to the temporal dependence of the transformation
$\hat{U}=\hat{U}\left(t\right)$.

Using Eqs.~(\ref{eq:Unit}), the transformed Hamiltonian acquires
a derivative with respect to the extended-space variable $\theta$:
\begin{equation}
\hat{K}\left(\theta,t\right)=-i\hbar\omega\frac{\partial}{\partial\theta}+\hat{H}\left(\theta,t\right),\label{eq:K-definition-explicit}
\end{equation}
In this way, the transformed Hamiltonian $\hat{K}\left(\theta,t\right)$
exhibits only a slow temporal dependence coming exclusively through
the second argument in $\hat{H}\left(\theta,t\right)$.

It is noteworthy that an equation of motion equivalent to Eq.~(\ref{eq:Shroedinger-transformed})
can also be obtained using a two-time ($t,t^{\prime}$) formalism~\cite{Howland1974,Breuer1989,Breuer1989a,Grifoni98,Drese1999,Eckardt08,Heinisch16}.
In the ($t,\theta^{\prime}$) notation, the formalism treats $\theta^{\prime}=\omega t+\theta$
and $t$ entering the time-dependent Hamiltonian $H\left(\theta^{\prime},t\right)$
as two independent variables. The temporal dependence of $\theta^{\prime}$
is then reflected by a derivative $\partial/\partial\theta^{\prime}$
which enters $\hat{K}\left(\theta^{\prime},t\right)$ defined in the
same manner as Eq.~(\ref{eq:K-definition-explicit}). At the end
of the calculations one recovers the physical solution by setting
$\theta^{\prime}=\omega t+\theta$. In the present formalism, this
operation corresponds to returning to the original state vector $\left|\phi_{\theta}\left(t\right)\right\rrangle =\left|\psi_{\theta+\omega t}\left(t\right)\right\rrangle $
via Eq.~(\ref{eq:state-vector-transformed}) involving the unitary
transformation $\hat{U}$ given by Eq.~(\ref{eq:U}).

\subsection{Hamiltonian in the abstract extended space}

It is convenient to characterize the basic vectors $e^{in\theta}$
only by a number $n$ without specifying the phase variable $\theta$.
For this let us introduce a set of abstract basis vectors $\left|\overline{n}\right\rangle $
corresponding to an orthogonal set of $\theta$-dependent functions:
$e^{in\theta}\leftrightarrow\left|\overline{n}\right\rangle $ with
$\left\langle \overline{m}\right.\left|\overline{n}\right\rangle =\delta_{nm}$.
In this representation (referred to as the abstract representation),
the original and transformed state vectors no longer include the angular
variable $\theta$ and can be cast in terms of $\left|\overline{n}\right\rangle $
as:\begin{subequations} \label{eq:abstract-space} 
\begin{align}
\left|\phi\left(t\right)\right\rrangle  & =\sum_{n=-\infty}^{\infty}\left|\phi^{(n)}\left(t\right)\right\rangle \left|\overline{n}\right\rangle ,\label{eq:abstract-space-phi}\\
\left|\psi\left(t\right)\right\rrangle  & =\sum_{n=-\infty}^{\infty}\left|\psi^{(n)}\left(t\right)\right\rangle \left|\overline{n}\right\rangle .\label{eq:abstract-space-psi}
\end{align}
\end{subequations}
On the other hand, the $\theta$-dependent extended-space
Hamiltonian $\hat{K}\left(\theta,t\right)$ is now replaced by an
abstract Hamiltonian $\hat{K}\left(t\right)$ given by 
\begin{equation}
\hat{K}\left(t\right)=\sum_{n=-\infty}^{\infty}\left|\overline{n}\right\rangle \hbar\omega n\left\langle \overline{n}\right|+\sum_{m,n=-\infty}^{\infty}H^{(m-n)}(t)\left|\overline{m}\right\rangle \left\langle \overline{n}\right|.\label{eq:K-expansion}
\end{equation}
In writing the first term of Eq.~(\ref{eq:K-expansion}) we noted
that $e^{in\theta}$ is an eigenfunction of the operator $-i\hbar\omega\partial/\partial\theta$
featured in Eq.~(\ref{eq:K-definition-explicit}) with an eigenvalue
$n\hbar\omega$. The second term contains the Fourier components $H^{\left(l\right)}(t)$
(with $l=m-n$) of the physical Hamiltonian given by Eq.~(\ref{eq:H-periodic-expansion}).
Here we used the fact that the exponents $e^{il\theta}$ entering
Eq.~(\ref{eq:H-periodic-expansion}) provide a shift of the abstract
state vectors: $\left|\overline{n}\right\rangle \rightarrow\left|\overline{n+l}\right\rangle $.

\begin{figure}
\begin{centering}
\begin{tabular}{c}
\includegraphics[width=0.99\columnwidth]{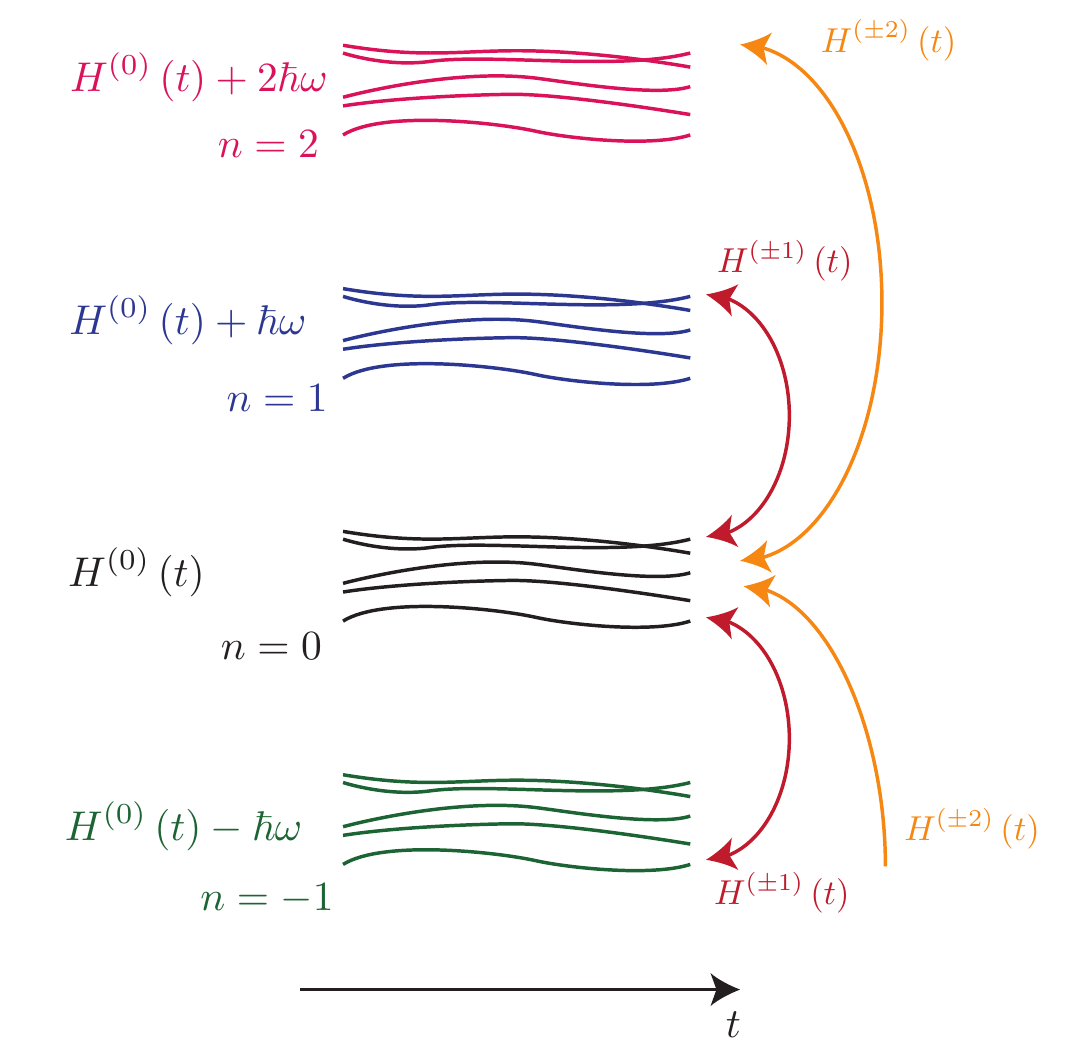} \tabularnewline
\end{tabular}
\par\end{centering}
\caption{\label{fig:Floquet-levels}Schematic visualization of the terms which enter
the slowly varying extended space Hamiltonian $\hat{K}=\hat{K}\left(t\right)$
given by Eqs.~(\ref{eq:K-expansion}) and (\ref{eq:K-expansion-1}).
Levels belonging to different Floquet bands $\left|\overline{n}\right\rangle $
are drawn in different colors. The operators $H^{\left(l\right)}(t)$
with $l\protect\ne0$ induce transitions between different Floquet
bands. The operators $H^{\left(0\right)}(t)-n\hbar\omega$ provide
the time-dependent zero-order eigenenergies of the extended-space
Hamiltonian. }
\end{figure}

The abstract extended-space Hamiltonian $\hat{K}=\hat{K}\left(t\right)$
can be represented as an infinite block matrix: 
\begin{widetext}
\begin{equation}
\hat{K}=\left(\begin{array}{cccccc}
\ldots & \ldots & \ldots & \ldots & \ldots & \ldots\\
\ldots & H^{\left(0\right)}-\hbar\omega & H^{\left(-1\right)} & H^{\left(-2\right)} & H^{\left(-3\right)} & \ldots\\
\ldots & H^{\left(1\right)} & H^{\left(0\right)} & H^{\left(-1\right)} & H^{\left(-2\right)} & \ldots\\
\ldots & H^{\left(2\right)} & H^{\left(1\right)} & H^{\left(0\right)}+\hbar\omega & H^{\left(-1\right)} & \ldots\\
\ldots & H^{\left(3\right)} & H^{\left(2\right)} & H^{\left(1\right)} & H^{\left(0\right)}+2\hbar\omega & \ldots\\
\ldots & H^{\left(4\right)} & H^{\left(3\right)} & H^{\left(2\right)} & H^{\left(1\right)} & \ldots\\
\ldots & \ldots & \ldots & \ldots & \ldots & \ldots
\end{array}\right),\label{eq:K-expansion-1}
\end{equation}
where matrix elements $K_{mn}=\left\langle \overline{m}\right|\hat{K}\left|\overline{n}\right\rangle =H^{(m-n)}+n\hbar\omega\delta_{n,m}$
are operators in the physical Hilbert space $\mathscr{H}$. The action
of the individual terms comprising the extended space Hamiltonian
is illustrated in Fig.~\ref{fig:Floquet-levels}. 
\end{widetext}

It is instructive that adding to $\hat{K}$ a unit operator $\hat{I}$
times $l\hbar\omega$ is equivalent to transforming $\hat{K}$ by
a unitary operator $\hat{P}_{l}$ 
\begin{equation}
\hat{K}+l\hbar\omega\hat{I}=\hat{P}_{l}^{\dagger}\hat{K}\hat{P}_{l}\label{eq:K-adding unit op-1}
\end{equation}
with 
\begin{equation}
\hat{P}_{l}=\sum_{n=-\infty}^{\infty}\left|\overline{n+l}\right\rangle \left\langle \overline{n}\right|.\label{eq:P_l}
\end{equation}
The operator $\hat{P}_{l}$ shifts an abstract state vector by $l$:
$\left|\overline{n}\right\rangle \rightarrow\left|\overline{n+l}\right\rangle $.
The relation~(\ref{eq:K-adding unit op-1}) implies that the spectrum
of $\hat{K}$ is invariant to shifting the energy by a multiple of
$\hbar\omega$. In fact, the operators $\hat{K}$ and $\hat{P}_{l}^{\dagger}\hat{K}\hat{P}_{l}$
are related by a unitary transformation and hence commute and have the same set of eigenstates. This property will be used in the subsequent
analysis of the high-frequency expansion of the Hamiltonian.

\subsection{Initial condition and subsequent evolution}

Let us consider a family of $\theta$-dependent state vectors $\left|\phi_{\theta}\left(t\right)\right\rangle $.
At the initial time $t=t_{0}$ the state vector must be chosen periodic
in $\theta$ according to Eq.~(\ref{eq:state-vector-periodic-cond}).
For convenience, we take a $\theta$-independent initial condition:
$\left|\phi_{\theta}\left(t_{0}\right)\right\rangle =\left|\alpha\right\rangle  $, where $\left|\alpha\right\rangle  $ is an initial state vector. In that case, the transformed state vector $\left|\psi_{\theta}\left(t\right)\right\rangle $ is also $\theta$-independent at the initial time $t=t_{0}$:
\begin{equation}
\left|\psi_{\theta}\left(t_{0}\right)\right\rangle =\left|\phi_{\theta-\omega t_{0}}\left(t_{0}\right)\right\rangle =\left|\alpha\right\rangle \quad\textrm{for all }\theta\in\left[0,2\pi\right).\label{eq:initial-condition-theta-independent}
\end{equation}
Therefore, initially one populates only the $n=0$ harmonic (in the
phase variable $\theta$) in the Fourier expansion of the original
or transformed state vector. In the abstract notation, the initial
state vector contains only the mode $\left|\overline{n}\right\rangle $
with $n=0$ in Eqs.~(\ref{eq:abstract-space}): 
\begin{equation}
\left|\phi\left(t_{0}\right)\right\rrangle =\left|\psi\left(t_{0}\right)\right\rrangle = \left|\alpha\right\rangle \left|\overline{0}\right\rangle  .\label{eq:initial-cond-abstract}
\end{equation}
Subsequently, for $t>t_{0}$ the state vector $\left|\psi_{\theta}\left(t\right)\right\rangle $
becomes $\theta$ dependent due to the $\theta$ dependence of the
Hamiltonian $\hat{K}\left(\theta,t\right)$ governing its temporal
evolution in Eq.~(\ref{eq:Shroedinger-transformed}). This means
the modes $\left|\overline{n}\right\rangle $ with $n\ne0$ appear
in the abstract state vector $\left|\psi\left(t\right)\right\rrangle $
during its subsequent time evolution described by the TDSE: 
\begin{equation}
i\hbar\partial_{t}\left|\psi\left(t\right)\right\rrangle =\hat{K}(t)\left|\psi\left(t\right)\right\rrangle .\label{eq:evolution-1}
\end{equation}
The dynamics governed by this equation will be analyzed in the next section.

\section{Temporal evolution and high frequency expansion\label{sec:Evolution and high frequency expansion}}

We shall make use of the symmetries of the Hamiltonian
$\hat{K}(t)$ for its block diagonalization. In doing so, we shall
include also the slow temporal dependence of $\hat{K}(t)$. We shall
concentrate on a high-frequency limit where $\omega$ exceeds all
other frequencies of the physical system. This will enable one to
find a high-frequency expansion of an effective Hamiltonian $H_{\textrm{eff}}\left(t\right)$
taking into account the slow temporal dependence of $\hat{K}(t)$.

\subsection{Block diagonalization}

We shall look for a unitary transformation 
\begin{equation}
\hat{D}^{\dagger}\left(t\right)\left|\psi\left(t\right)\right\rrangle =\left|\chi\left(t\right)\right\rrangle \label{eq:D-transfomed-state-vector}
\end{equation}
which leads to a TDSE

\begin{equation}
i\hbar\partial_{t}\left|\chi\left(t\right)\right\rrangle =\hat{K}_{D}(t)\left|\chi\left(t\right)\right\rrangle \label{eq:TDSE-Block-diagonal}
\end{equation}
governed by a block-diagonal (in the extended space) Hamiltonian 
\begin{equation}
\begin{aligned} & \hat{K}_{D}\left(t\right)=\hat{D}^{\dagger}\left(t\right)\hat{K}\left(t\right)\hat{D}\left(t\right)-i\hbar\hat{D}^{\dagger}\left(t\right)\dot{\hat{D}}\left(t\right)\\
 & =\sum_{n}\left|\overline{n}\right\rangle \left(H_{\textrm{eff}}\left(t\right)+n\hbar\omega\right)\left\langle \overline{n}\right|\,.
\end{aligned}
\label{eq:Block-diag}
\end{equation}
Here $H_{\textrm{eff}}\left(t\right)$ entering the block-diagonal
operator $\hat{K}_{D}\left(t\right)$ represents a slowly varying
Floquet Hamiltonian describing an effective evolution of the physical
system. It is instructive that the transformed Hamiltonian $\hat{K}_{D}\left(t\right)$
contains an additional term $i\hbar\hat{D}^{\dagger}\left(t\right)\dot{\hat{D}}\left(t\right)$
due to the temporal dependence of the unitary transformation $\hat{D}\left(t\right)$.
Therefore, the block diagonalization is to be carried out in a self-consistent manner with respect to $\hat{D}\left(t\right)$.

Denoting 
\begin{equation}
\hat{N}=\sum_{n=-\infty}^{\infty}\left|\overline{n}\right\rangle n\left\langle \overline{n}\right|\,\quad\mathrm{and}\quad\hat{I}=\sum_{n=-\infty}^{\infty}\left|\overline{n}\right\rangle \left\langle \overline{n}\right|\,,\label{eq:N}
\end{equation}
one can write 
\begin{equation}
\hat{K}_{D}\left(t\right)=\hbar\omega\hat{N}+H_{\textrm{eff}}\left(t\right)\hat{I}\,.\label{eq:Block-diag-compact}
\end{equation}

It is noteworthy that $H_{\textrm{eff}}\left(t\right)$ is not necessarily
diagonal in the physical space. Furthermore, the block diagonalization
is not a unique procedure. It is defined only up to a unitary transformation
in the physical space, the same for each block comprising $\hat{K}_{D}$
in Eqs.~(\ref{eq:Block-diag})--(\ref{eq:Block-diag-compact}). However,
performing a high-frequency expansion in the powers of $1/\omega$,
the block-diagonal operator $\hat{K}_{D}$ becomes unique provided
the zero-order term of the diagonalization operator $\hat{D}$ is
set to a unit operator, $\hat{D}_{\left(0\right)}=\hat{I}$. In fact,
in the limit of an infinite frequency, $H^{\left(l\right)}/\omega\rightarrow0$,
the off-diagonal elements $H^{\left(l\right)}$ (with $l\ne0$) of
the extended-space Hamiltonian $\hat{K}$ can be neglected by replacing
$\hat{K}\rightarrow\hbar\omega\hat{N}+H^{\left(0\right)}\hat{I}\,$.
This means that for $H^{\left(l\right)}/\omega\rightarrow0$ one can
take: $\hat{D}=\hat{D}_{\left(0\right)}=\hat{I}$.

Note also that a block diagonalization similar to that in Eq.~(\ref{eq:Block-diag})
has been employed in Ref.~\cite{Eckardt2015} when dealing with a
high-frequency expansion of the effective Hamiltonian $H_{\textrm{eff}}$
(see also Refs.~\cite{Goldman2014,Eckardt2015,Bukov2015,Mikami16PRB,Eckardt16-Review}
). However, in the present situation the extended-space Hamiltonian
$\hat{K}\left(t\right)$ additionally depends on a slow time, so the
effective Hamiltonian $H_{\textrm{eff}}\left(t\right)$ and the diagonalization
operator $\hat{D}$ are also time dependent. Furthermore, an extra
term $i\hbar\hat{D}^{\dagger}\dot{\hat{D}}$ in Eq.~(\ref{eq:Block-diag})
provides additional contributions to the high-frequency expansions
of these operators due to the temporal changes of the components $H^{\left(l\right)}\left(t\right)$
entering the physical Hamiltonian~(\ref{eq:H-periodic-expansion}),
as we shall see in Sec.~\ref{sec:High-frequency-expansion}.

In this way, our formalism combines two approaches: a systematic high-frequency
expansion of $H_{\textrm{eff}}\left(t\right)$ via a degenerate perturbation
theory in the extended Floquet space~\cite{Eckardt2015}, as well
as an adiabatic perturbation theory~\cite{Drese1999,Weinberg16arXiv}
with respect to the basis vectors $\left|\overline{n}\right\rangle $
of the subspace $\mathscr{T}$ due to the temporal changes of the
block-diagonalization operator $\hat{D}(t)$.

The relation~(\ref{eq:K-adding unit op-1}) implies that the unitary
operator $\hat{D}$ diagonalizing $\hat{K}$ is invariant with respect
to the shift operator: $\hat{D}\left(t\right)=\hat{P}_{l}^{\dagger}\hat{D}\left(t\right)\hat{P}{}_{l}$.
This is the case if $\hat{D}$ has the following form~\cite{Eckardt2015}:
\begin{equation}
\hat{D}\left(t\right)=\sum_{m,n=-\infty}^{\infty}D^{(m-n)}\left(t\right)\left|\overline{m}\right\rangle \left\langle \overline{n}\right|=\sum_{l=-\infty}^{\infty}D^{(l)}\left(t\right)\hat{P}_{l}\,.\label{eq:D-expansion}
\end{equation}

\subsection{Temporal evolution in the extended space}

Since $\hat{K}_{D}\left(t\right)$ entering the TDSE~(\ref{eq:TDSE-Block-diagonal})
is block diagonal, the temporal evolution of the transformed state
vector $\left|\chi\left(t\right)\right\rrangle =\sum_{n}\left|\chi_{n}\left(t\right)\right\rangle \left|\overline{n}\right\rangle $
is described by a set of uncoupled TDSEs for the constituting state vectors
$\left|\chi_{n}\left(t\right)\right\rangle $ governed by the Hamiltonians
$H_{\textrm{eff}}\left(t\right)+n\hbar\omega$. As a result, the time evolution
of the transformed state vector $\left|\chi\left(t\right)\right\rrangle $
is given by 
\begin{equation}
\left|\chi\left(t\right)\right\rrangle =\sum_{n=-\infty}^{\infty}\exp\left[-in\omega\left(t-t_{0}\right)\right]\left|\overline{n}\right\rangle U_{\textrm{eff}}\left(t,t_{0}\right)\left|\chi_{n}\left(t_{0}\right)\right\rangle \,,\label{eq:chi-formal-solution}
\end{equation}
where a unitary operator $U_{\textrm{eff}}\left(t,t_{0}\right)$ describes
a quantum evolution in the physical space $\mathscr{H}$ generated
by the effective Hamiltonian $H_{\textrm{eff}}\left(t\right)$: 
\begin{equation}
i\hbar\partial_{t}U_{\textrm{eff}}\left(t,t_{0}\right)=H_{\textrm{eff}}\left(t\right)U_{\textrm{eff}}\left(t,t_{0}\right)\,,\quad U_{\textrm{eff}}\left(t_{0},t_{0}\right)=I. \label{eq:TDSE_effect-U_eff}
\end{equation}
If the effective Hamiltonian does not commute with itself at different
times $\left[H_{\textrm{eff}}\left(t^{\prime}\right),H_{\textrm{eff}}\left(t^{\prime\prime}\right)\right]\ne0$,
a formal solution to this equation involves a time ordering 
\begin{equation}
U_{\textrm{eff}}\left(t,t_{0}\right)={\cal T}\exp\left[-\frac{i}{\hbar}\int_{t_{0}}^{t}H_{\textrm{eff}}\left(t^{\prime}\right)\mathrm{d}t^{\prime}\right]\,.\label{U_eff}
\end{equation}
For sufficiently slow changes of $H_{\textrm{eff}}\left(t^{\prime}\right)$
the adiabatic approximation~\cite{Berry:1984,Mead1992} can be applied
to find the evolution operator $U_{\textrm{eff}}\left(t,t_{0}\right)$
on the basis of instantaneous eigenstates of $H_{\textrm{eff}}\left(t^{\prime}\right)$.

Combining Eqs.~(\ref{eq:initial-cond-abstract}) and~(\ref{eq:D-transfomed-state-vector}),
the initial condition for the transformed state vector reads as $\left|\chi\left(t_{0}\right)\right\rrangle =\hat{D}^{\dagger}\left(t_{0}\right)\left|\overline{0}\right\rangle \left|\alpha\right\rangle $,
giving 
\begin{equation}
\left|\chi_{n}\left(t_{0}\right)\right\rangle =D^{\left(-n\right)\dagger}\left(t_{0}\right)\left|\alpha\right\rangle, \label{eq:chi-t_0}
\end{equation}
where $D^{(m-n)}=\left\langle \overline{m}\right|\hat{D}\left(t\right)\left|\overline{n}\right\rangle $
is an operator acting in the physical space $\mathscr{H}$. Substituting
Eq.~(\ref{eq:chi-t_0}) into~(\ref{eq:chi-formal-solution}), the
extended-space state vector $\left|\psi\left(t\right)\right\rrangle =\hat{D}\left(t\right)\left|\chi\left(t\right)\right\rrangle $
can be expressed in terms of the initial state vector $\left|\alpha\right\rangle $
and the matrix elements of the transformation operator $\hat{D}\left(t\right)$:

\begin{equation}
\begin{aligned}\left|\psi\left(t\right)\right\rrangle =\sum_{n,l=-\infty}^{\infty} & \left|\overline{l}\right\rangle D^{\left(l-n\right)}\left(t\right)e^{-in\omega\left(t-t_{0}\right)}\\
 & \times U_{\textrm{eff}}\left(t,t_{0}\right)D^{\left(-n\right)\dagger}\left(t_{0}\right)\left|\alpha\right\rangle \,.
\end{aligned}
\label{eq:psi-t-solution}
\end{equation}

\subsection{Temporal evolution in the physical space}

Transition to the $\theta$ representation $\left|\psi\left(t\right)\right\rrangle \rightarrow\left|\psi_{\theta}\left(t\right)\right\rrangle $
is carried out replacing $\left|\overline{l}\right\rangle \rightarrow e^{il\theta}$
in Eq.~(\ref{eq:psi-t-solution}). Using Eq.~(\ref{eq:state-vector-transformed}),
one arrives at the extended-space state vector in the original representation $\left|\phi_{\theta}\left(t\right)\right\rrangle =\left|\psi_{\theta+\omega t}\left(t\right)\right\rrangle $.
It can be treated as a state vector $\left|\phi_{\theta}\left(t\right)\right\rangle$ of the physical space $\mathscr{H}$ exhibiting a parametric dependence on the phase $\theta$ featured in the Hamiltonian $H\left(\theta+\omega t,t\right)$: 
\begin{equation}
\begin{aligned} & \left|\phi_{\theta}\left(t\right)\right\rangle \\
 & =U_{\textrm{Micro}}\left(\omega t+\theta,t\right)U_{\textrm{eff}}\left(t,t_{0}\right)U_{\textrm{Micro}}^{\dagger}\left(\omega t_{0}+\theta,t_{0}\right)\left|\alpha\right\rangle \,,
\end{aligned}
\label{eq:phi-t-solution}
\end{equation}
where 
\begin{equation}
U_{\textrm{Micro}}\left(\omega t+\theta,t\right)=\sum_{n=-\infty}^{\infty}D^{\left(n\right)}\left(t\right)e^{in\left(\omega t+\theta\right)}\,.\label{eq:U_Kick}
\end{equation}
is a unitary operator describing the micromotion (see Appendix~\ref{sec:Appendix-A:-Unitarity U}).
It can be cast in the exponential form $U_{\textrm{Micro}}=\exp\left(-iS_{\textrm{Micro}}\right)$,
where a Hermitian operator $S_{\textrm{Micro}}$ featured in the exponent
is usually referred to as a micromotion operator (known also as a
kick operator)~\cite{Rahav03,Eckardt2015,Goldman2014,Bukov2015,Eckardt16-Review}.
We will use the term ``micromotion operator'' also for the unitary
operator $U_{\textrm{Micro}}$.

Equation~(\ref{eq:phi-t-solution}) represents a generalization of
the Floquet theorem to periodically modulated Hamiltonians $H\left(\omega t+\theta,t\right)$
containing an extra temporal dependence. The dynamics of the system
is then described in an effective manner by the slowly varying
Hamiltonian $H_{\textrm{eff}}=H_{\textrm{eff}}\left(t\right)$
via the unitary operator $U_{\textrm{eff}}\left(t,t_{0}\right)$ defined
by Eqs.~(\ref{eq:TDSE_effect-U_eff}) and (\ref{U_eff})\footnote{Previously the notion of time-dependent effective Hamiltonians was used in the Supplementary material of Ref.~\cite{nascimbene15}  to describe a slow ramp of a dynamical optical lattice with a sub-wavelength spacing. The ramp was split into a set of stroboscopic pieces, in which the modulation was assumed to be constant in amplitude.
In such an approach, the obtained time-dependent effective Hamiltonian contains a contribution that depends on the initial phase of the drive. However, this should not be considered as a genuine contribution to the effective Hamiltonian. The latter $H_{\textrm{eff}}\left(t\right)$ is associated with a long-time dynamics and should be independent of the initial phase, as in the case of the stationary driving \cite{Rahav03,Goldman2014,Eckardt2015}.}.
Additionally the solution~(\ref{eq:phi-t-solution}) contains the
micromotion operator $U_{\textrm{Micro}}\left(\omega t+\theta,t\right)$
calculated at the initial and final times $t=t_{0}$ and $t$.

It is instructive that the effective Hamiltonian $H_{\textrm{eff}}\left(t\right)$
and hence the unitary operator $U_{\textrm{eff}}\left(t,t_{0}\right)$
describing an effective long time evolution in Eq.~(\ref{eq:phi-t-solution})
do not depend on the phase $\theta$ entering the original Hamiltonian
$H\left(\theta+\omega t,t\right)$. Only the micromotion operators
$U_{\textrm{Micro}}\left(\omega t+\theta,t\right)$ and $S_{\textrm{Micro}}\left(\omega t+\theta,t\right)$
are $\theta$ dependent. 
Yet, in comparison
to the previous studies~\cite{Eckardt2015,Goldman2014,Bukov2015,Eckardt16-Review},
the operator $U_{\textrm{Micro}}\left(\omega t+\theta,t\right)$ includes
not only the fast micromotion represented by the exponential factors
$e^{in\left(\omega t+\theta\right)}$ in Eq.~(\ref{eq:U_Kick}),
but also an additional temporal dependence due to slow changes of
the transformation $\hat{D}\left(t\right)$ diagonalizing the extended-space Floquet Hamiltonian $\hat{K}\left(t\right)$ in Eq.~(\ref{eq:Block-diag}).
In particular, this is the case if the periodic perturbation is switched
on and off in a smooth manner, which is relevant to, e.g., shaken optical
lattices with a time-dependent shaking strength~\cite{Eckardt2010,Aidelsburger:2011,Arimondo2012,Hauke:2012,Windpassinger2013RPP,Struck:2013,Aidelsburger:2013,Jotzu2014,Aidelsburger14NP,Kennedy15,Flaschner16}.
In that case, the micromotion operator featured in Eq.~(\ref{eq:phi-t-solution}) reduces to the unit operator 
\begin{equation}
\left. U_{\textrm{Micro}}^{\dagger}\left(\omega t+\theta,t\right)\right|_{t=t_0}=1
\label{eq:micro_1}
\end{equation}
at the initial time $t_{0}$, when the periodic perturbation starts slowly switching on. In that case, the extended-space Hamiltonian $\hat{K}(t)$ given by Eq.~(\ref{eq:K-expansion-1}) is block diagonal at the initial time $t=t_0$ when the periodic perturbation is off, so no subsequent block diagonalization is needed.   

\subsection{High-frequency expansion\label{sec:High-frequency-expansion}}

Knowing the effective Hamiltonian $H_{\textrm{eff}}\left(t\right)$
and the micromotion operator $U_{\textrm{Micro}}\left(\omega t+\theta,t\right)$ one
can use Eqs.~(\ref{eq:phi-t-solution}) and~(\ref{eq:TDSE_effect-U_eff}) and~(\ref{U_eff})
to find the time evolution of the state vector $\left|\phi_{\theta}\left(t\right)\right\rangle $
which parametrically depends on the phase $\theta$. Usually, both
operators $H_{\textrm{eff}}\left(t\right)$ and $U_{\textrm{Micro}}\left(\omega t+\theta,t\right)$
can not be determined analytically. However, for sufficiently high driving
frequencies, they can be expressed as a series expansion in the terms
of the powers of $\omega^{-1}$. This can be done if off-diagonal
matrix elements of the extended-space Hamiltonian~(\ref{eq:K-expansion-1})
are small compared to the driving frequency, $\left|H_{\alpha\beta}^{\left(l\right)}\right|\ll\hbar\omega$,
and the spectral width of the physical system is much smaller than
driving frequency, $\left|\varepsilon_{\alpha}-\varepsilon_{\beta}\right|=\left|H_{\alpha\alpha}^{\left(0\right)}-H_{\beta\beta}^{\left(0\right)}\right|\ll\hbar\omega$.
Furthermore, the operators $H^{\left(l\right)}$ should change little
over a period of oscillations: $\left|\dot{H}_{\alpha\beta}^{\left(l\right)}\right|\ll\omega\left|H_{\alpha\beta}^{\left(l\right)}\right|$.
The latter condition appears because the matrix elements of the operator
$i\hbar\hat{D}^{\dagger}\left(t\right)\dot{\hat{D}}\left(t\right)$ featured
in Eq.~(\ref{eq:Block-diag}) should be much smaller than $\omega$.

In some cases, such as in shaken optical lattices~\cite{Eckardt2010,Aidelsburger:2011,Arimondo2012,Hauke:2012,Windpassinger2013RPP,Struck:2013,Aidelsburger:2013,Jotzu2014,Aidelsburger14NP,Kennedy15,Flaschner16},
the spectrum of the physical system extends beyond the driving frequency,
so the condition $\left|\varepsilon_{\alpha}-\varepsilon_{\beta}\right|\ll\hbar\omega$
does not hold for the states with high energies $\varepsilon_{\alpha}$.
Yet, if these states are not directly accessible from the initial state
of the system, the high-frequency expansion can still be used to describe
the dynamics of the system at the intermediate times when the higher
states are not yet populated~\cite{Eckardt16-Review}. In particular,
it was demonstrated \cite{kuwahara16} that for \emph{time-periodic}
systems the truncated high-frequency expansion can remain applicable
even when the condition $\left|\varepsilon_{\alpha}-\varepsilon_{\beta}\right|\ll\hbar\omega$
is not met. On the other hand, in many-body systems the adiabatic
approximation may break down not only at very high ramp rates, but
also at very slow ones due to avoided crossings of Floquet many-body
resonances~\cite{Eckardt08,Weinberg16arXiv,Eckardt16-Review}
\cite{[{This kind of breakdown can occur not only in many-body systems, 
but also for single-particle systems, such as a driven anharmonic oscillator 
or a particle in a square potential considered by }] [{. These systems do not 
posses an adiabatic limit in the usual sense due to the denseness of the quasienergy 
spectrum.}]Hone97}. This effect is not captured by the high-frequency expansion, but
it should become smaller and smaller with increasing the ramp rates
and driving frequency.

A general formalism of the high-frequency expansion is presented in
Appendix~\ref{sec:Appendix-B:-Expansion}. Here, we summarize
the findings. 
\begin{widetext}
The effective Hamiltonian expanded in the powers of $\omega^{-1}$
reads as
\begin{equation}
H_{\textrm{eff}}=H_{\textrm{eff}\left(0\right)}+H_{\textrm{eff}\left(1\right)}+H_{\textrm{eff}\left(2\right)}+\ldots, \label{eq:H_eff_expansion}
\end{equation}
where the $n$th term $H_{\textrm{eff}\left(n\right)}$ is proportional
to $\omega^{-n}$. The first three expansion terms are
\begin{subequations}
\label{eq:H_eff} 
\begin{align}
H_{\textrm{eff}\left(0\right)} & =H^{\left(0\right)},\label{eq:H_eff_0}\\
H_{\textrm{eff}\left(1\right)} & =\frac{1}{\hbar\omega}\sum_{m=1}^{\infty}\frac{1}{m}\left[H^{\left(m\right)},H^{\left(-m\right)}\right],\label{eq:H_eff_1}\\
H_{\textrm{eff}\left(2\right)} & =\frac{1}{\left(\hbar\omega\right)^{2}}\sum_{m\neq0}\left\{ \frac{\left[H^{\left(-m\right)},\left[H^{\left(0\right)},H^{\left(m\right)}\right]\right]-i\hbar\left[H^{\left(-m\right)},\dot{H}^{\left(m\right)}\right]}{2m^{2}}+\sum_{n\neq\left\{0, m\right\}}\frac{\left[H^{\left(-m\right)},\left[H^{\left(m-n\right)},H^{\left(n\right)}\right]\right]}{3mn}\right\} ,\label{eq:H_eff_2}
\end{align}
\end{subequations}
where the (slow) temporal dependence of the components
$H^{\left(n\right)}=H^{\left(n\right)}\left(t\right)$ is kept implicit.

Note that the second-order contribution proportional to $\left[H^{\left(-m\right)},\dot{H}^{\left(m\right)}\right]$
stems from projecting onto a selected Floquet band (with $n=0$) of
an extra term $-i\hbar\hat{D}^{\dagger}\left(t\right)\dot{\hat{D}}\left(t\right)$
entering Eq.~(\ref{eq:Block-diag}). This provides a geometric phase~\cite{Berry:1984}
for an adiabatic motion in the selected Floquet band. The geometric
phase can be non-Abelian if more than one quantum state is involved
in the adiabatic motion~\cite{Wilczek:1984,Moody1986,Zee1988}. In
Sec.~\ref{sec:Oscillating spin} we shall consider an example
providing non-Abelian geometric phases for the adiabatic motion in
the Floquet band.

Expanding the Hermitian micromotion operator $S_{\textrm{Micro}}$
entering $U_{\textrm{Micro}}=\exp\left(-iS_{\textrm{Micro}}\right)$
in the series $S_{\textrm{Micro}}=S_{\textrm{Micro}\left(1\right)}+S_{\textrm{Micro}\left(2\right)}+\ldots$,
the first- and the second-order terms read as
\begin{subequations} \label{eq:S}
\begin{equation}
S_{\textrm{Micro}\left(1\right)}\left(\theta^{\prime},t\right)=\frac{1}{i\hbar\omega}\sum_{m\ne0}\frac{1}{m}H^{\left(m\right)}e^{im\theta^{\prime}},\label{eq:S_1}
\end{equation}
\begin{equation}
S_{\textrm{Micro}\left(2\right)}\left(\theta^{\prime},t\right)=\frac{1}{2i\left(\hbar\omega\right)^{2}}\sum_{m\ne0}\left\{ \frac{1}{m^{2}}\left[H^{\left(m\right)},H^{\left(0\right)}\right]+\sum_{n\neq0}\frac{1}{mn}\left[H^{\left(n\right)},H^{\left(m-n\right)}\right]+\frac{2i\hbar}{m^{2}}\dot{H}^{\left(m\right)}\right\} e^{im\theta^{\prime}},\label{eq:S_2}
\end{equation}
\end{subequations}
where $\theta^{\prime}=\omega t+\theta$.

On the other hand, the expansion of the operator $U_{\textrm{Micro}}\left(\theta^{\prime},t\right)$
up to the $\omega^{-3}$ order is given by 
\begin{align}
U_{\textrm{Micro}}\left(\theta^{\prime},t\right) & =\mathbf{1}_{\mathscr{H}}-\frac{1}{\hbar\omega}\sum_{m\ne0}\frac{1}{m}H^{\left(m\right)}e^{im\theta^{\prime}}+\frac{1}{2\left(\hbar\omega\right)^{2}}\sum_{m\neq0}\sum_{n\neq0}e^{i\left(m+n\right)\theta^{\prime}}\frac{H^{\left(m\right)}H^{\left(n\right)}}{nm}\nonumber \\
 & +\frac{1}{2\left(\hbar\omega\right)^{2}}\sum_{m\ne0}e^{im\theta^{\prime}}\left\{ \frac{\left[H^{\left(0\right)},H^{\left(m\right)}\right]-2i\hbar\dot{H}^{\left(m\right)}}{m^{2}}-\sum_{n\neq0}\frac{\left[H^{\left(n\right)},H^{\left(m-n\right)}\right]}{nm}\right\} +\mathcal{O}\left(\omega^{-3}\right)\,.\label{eq:U_kick_expansion}
\end{align}
Although the operator $U_{\textrm{Micro}}$ is unitary, it becomes
non-unitary if approximated with a finite number of terms. For instance,
in Eq.~(\ref{eq:U_kick_expansion}) the unitarity holds up to the
$\omega^{-3}$ order. 
\end{widetext}

Generalizing Refs.~\cite{Eckardt2015,Goldman2014,Bukov2015,Mikami16PRB,Eckardt16-Review},
the expanded effective Hamiltonian and micromotion operators are now
time dependent due to the temporal dependence of the components $H^{\left(m\right)}=H^{\left(m\right)}\left(t\right)$
entering the expansions. For instance, if the amplitude of the periodic
perturbation applied to the system slowly increases from zero reaching
a saturation value at $t$, the operator $U_{\textrm{Micro}}(t)$
is the unit operator at $t=t_{0}$, and reaches a stationary oscillating
solution at the saturation times $t$. Furthermore, in the present
situation the effective Hamiltonian and micromotion operators acquire additional terms due to the slow 
temporal dependence of the harmonics $H^{\left(m\right)}\left(t\right)$. Specifically, the term proportional to 
$\left[H^{\left(-m\right)},\dot{H}^{\left(m\right)}\right]$ appears as the second-order correction to the effective Hamiltonian in Eq.~(\ref{eq:H_eff_2}).
On the other hand, the terms proportional  to $\dot{H}^{\left(m\right)}$ represent the first-order correction to the micromotion operators in 
Eqs.~(\ref{eq:S_2}) and (\ref{eq:U_kick_expansion}).

The high-frequency expansion of the Hamiltonian is often restricted to the  zero and first 
orders, in which the extra term  $\propto \left[H^{\left(-m\right)},\dot{H}^{\left(m\right)}\right]$ 
does not show up. In that case one can simply replace the time-independent effective Hamiltonian obtained for the stationary driving by the time dependent one. For example, shaking of optical lattices is known to renormalize inter-site tunneling amplitudes~\cite{eckardt05,lignier07,Arimondo2012,Bukov2015,Eckardt16-Review} which acquire a slow temporal dependence in the case of a slowly 
varying driving. In Appendix~\ref{sec:1d-lattice}, this is illustrated for a one-dimensional shaken optical lattice with a slowly changing amplitude of driving.

In the following Sec.~\ref{sec:Oscillating spin} we will consider a spin in an oscillating 
magnetic field with a changing direction. In that case, there are no zero- and first-order 
contributions to the effective Hamiltonian. Therefore, the second-order term
proportional to $\left[H^{\left(-m\right)},\dot{H}^{\left(m\right)}\right]$ represents 
a dominant contribution which plays a vital role in the system
dynamics providing non-Abelian geometric phases.

\section{Spin in an oscillating magnetic field\label{sec:Oscillating spin}}

Let us apply the general formalism to a spin in a fast oscillating
magnetic field $\mathbf{B}\left(t\right)\cos\left(\omega t+\theta\right)$
with a slowly varying amplitude $\mathbf{B}\left(t\right)$. Such a system is described by a Hamiltonian

\begin{equation}
H\left(\omega t+\theta,t\right)=g_{F}\mathbf{F}\cdot\mathbf{B}\left(t\right)\cos\left(\omega t+\theta\right)\,,\label{eq:H-spin}
\end{equation}
where $g_{F}$ is a gyromagnetic factor, $\mathbf{F}=F_{1}\mathbf{e}_{x}+F_{2}\mathbf{e}_{y}+F_{3}\mathbf{e}_{z}$
is a spin operator satisfying the usual commutation relations $\left[F_{l},F_{m}\right]=i\hbar\epsilon_{lmn}F_{n}$.
Here, $\epsilon_{lmn}$ is a Levi-Civita symbol, and a summation over
a repeated Cartesian index $n=1,2,3$ is implied. The non-zero Fourier
components of the Hamiltonian~(\ref{eq:H-spin}) are 
\begin{equation}
H^{\left(1\right)}=H^{\left(-1\right)}=\frac{g_{F}}{2}\mathbf{F}\cdot\mathbf{B}\left(t\right).
\end{equation}

We now obtain the effective Hamiltonian and the micromotion operators up to the second order in $\omega^{-1}$ inclusively. Calling on Eqs.~(\ref{eq:H_eff_expansion}) and (\ref{eq:H_eff}), the truncated effective Hamiltonian reads as: 
\begin{equation}
\begin{aligned}H_{\textrm{eff}} & =H_{\textrm{eff}\left(2\right)}=\frac{-i\hbar}{\left(\hbar\omega\right)^{2}}\left[H^{\left(1\right)},\dot{H}^{\left(1\right)}\right]\\
 & =\frac{-i\hbar g_{F}^{2}}{\left(2\hbar\omega\right)^{2}}B_{k}\dot{B}_{l}\left[F_{k},F_{l}\right]\\
 & =g_{F}^{2}\left(2\omega\right)^{-2}\epsilon_{klm}B_{k}\dot{B}_{l}F_{m}\\
 & =g_{F}^{2}\left(2\omega\right)^{-2}\mathbf{F}\cdot\left(\mathbf{B}\times\dot{\mathbf{B}}\right).
\end{aligned}
\label{eq:spin_effect}
\end{equation}
Using Eq.~(\ref{eq:S_1}), the first-order micromotion operator is
given by:

\begin{equation}
S_{\textrm{Micro}(1)}\left(\omega t+\theta,t\right)=\frac{g_{F}}{\hbar\omega}\mathbf{F}\cdot\mathbf{B}\left(t\right)\sin\left(\omega t+\theta\right)\,.\label{eq:U_Micro-spin-1,2}
\end{equation}
The second-order contribution to the micromotion given by~(\ref{eq:S_2})
appears now due to ramping of the magnetic field:

\begin{equation}
S_{\textrm{Micro}\left(2\right)}\left(\omega t+\theta,t\right)=\frac{g_{F}}{\hbar\omega^{2}}\mathbf{F}\cdot\dot{\mathbf{B}}\left(t\right)\cos\left(\omega t+\theta\right).\label{eq:U_Micto-spin-2-time-dep}
\end{equation}

According to Eq.~(\ref{eq:spin_effect}), the change in the orientation
of the magnetic field provides an effective Hamiltonian $H_{\textrm{eff}}$
proportional to the spin perpendicular to both the magnetic field
$\mathbf{B}$ and its derivative $\dot{\mathbf{B}}$, i.e., perpendicular
to the rotation plane for the magnetic field. If the plane of the
rotation is changing, the Hamiltonian does not commute with itself
at different times, so the time ordering is needed in the evolution
operator $U_{\textrm{eff}}\left(t,t_{0}\right)$ presented in Eq.~(\ref{U_eff-spin})
below. The effective evolution of the spin is then associated with
non-Abelian (noncommuting) geometric phases, as we shall see below.

It is to be emphasized that in the present situation the geometric
phases appear because the effective evolution of the physical system
involves the adiabatic elimination of the Floquet bands with $m\ne0$
in the extended space, as generally illustrated in Fig.~\ref{fig:Floquet-levels}.
Thus, the emerging non-Abelian phases reflect the geometry of
the extended Floquet space rather than that of the physical one.

To see the geometric nature of the effective Hamiltonian~(\ref{eq:spin_effect}),
it is convenient to represent it in terms of a geometric vector potential
$\mathbf{{\cal A}}$:

\begin{equation}
H_{\textrm{eff}}=\mathbf{{\cal A}}\cdot\dot{\mathbf{B}}\,,\qquad\mathbf{{\cal A}}=g_{F}^{2}\left(2\omega\right)^{-2}\left(\mathbf{F}\times\mathbf{B}\right)\,.\label{eq:H_eff-spin-A}
\end{equation}
The evolution operator~(\ref{U_eff}) then takes the form 
\begin{equation}
U_{\textrm{eff}}\left(t,t_{0}\right)={\cal T}\exp\left[-\frac{i}{\hbar}\int_{t_{0}}^{t}\mathbf{{\cal A}}\cdot d\mathbf{B}\left(t^{\prime}\right)\right]\,.\label{U_eff-spin}
\end{equation}
The operator $U_{\textrm{eff}}\left(t,t_{0}\right)$ is thus determined
by a path of the magnetic field, not by a speed of its change, showing
a geometric origin of the acquired phases.

In particular, performing an anticlockwise rotation of the magnetic
field $\mathbf{B}$ by an angle $\varphi$ in a plane orthogonal to
a unit vector $\mathbf{n}\propto\mathbf{B}\times\dot{\mathbf{B}}$,
the corresponding evolution operator is defined by a spin along the
rotation direction: $\mathbf{F}\cdot\mathbf{n}$. If additionally
an amplitude $B$ of the rotating magnetic field is not changing,
the evolution operator (\ref{U_eff-spin}) simplifies to 
\begin{equation}
U_{\textrm{eff}}\left(\mathbf{n},\varphi\right)=\exp\left[-\frac{i}{\hbar}\gamma_{\varphi}\mathbf{F}\cdot\mathbf{n}\right]\,,\qquad\gamma_{\varphi}=\varphi\frac{g_{F}^{2}B^{2}}{4\omega^{2}}\,.\label{U_eff-spin-2}
\end{equation}

After making $l$ rotations, the angle is given by $\varphi=2\pi l$,
where $l$ is an integer. In that case, the magnetic field $\mathbf{B}\left(t\right)$ comes back
to its original value. Therefore, the exponent $\gamma_{\varphi}\mathbf{F}\cdot\mathbf{n}/\hbar$
can be identified as a Wilczek-Zee phase operator~\cite{Wilczek:1984}
representing a non-Abelian generalization to the Berry phase~\cite{Berry:1984}.
The corresponding eigenvalues $\gamma_{\varphi}m_{F}$ linearly depend
on the spin projection $m_{F}$ along the rotation axis $\mathbf{n}$.
For a single loop ($l=1$) the phase $\gamma_{\varphi}$ is much smaller
than the unity because of the assumption of the high-frequency driving. Performing
many loops ($l\gg1$), one may accumulate a considerable phase $\gamma_{\varphi}$. 
It is noteworthy that two consecutive rotations along non-parallel
axes $\mathbf{n}^{\prime}$ and $\mathbf{n}^{\prime\prime}$ do not
commute $\left[U_{\textrm{eff}}\left(\mathbf{n}^{\prime},\varphi^{\prime}\right),U_{\textrm{eff}}\left(\mathbf{n}^{\prime\prime},\varphi^{\prime\prime}\right)\right]\ne0$.
This demonstrates a non-Abelian character of the problem.

As shown in Appendix~\ref{sec:Spin--Relation to rotation frequency shift},
the acquired geometric phase $\gamma_{\varphi}\mathbf{F}\cdot\mathbf{n}/\hbar$ 
entering the evolution operator (\ref{U_eff-spin-2}) stems from the rotational frequency shift~\cite{birula97rfs}
representing a correction to it. The correction term to the effective Hamiltonian presented by
Eq.~(\ref{eq:rot_eff_2}) is proportional to the spin along the rotation
direction, in agreement with  Eqs.~(\ref{eq:spin_effect}) and (\ref{U_eff-spin-2}). 

\begin{figure}
\begin{centering}
\begin{tabular}{c}
\includegraphics[width=0.99\columnwidth]{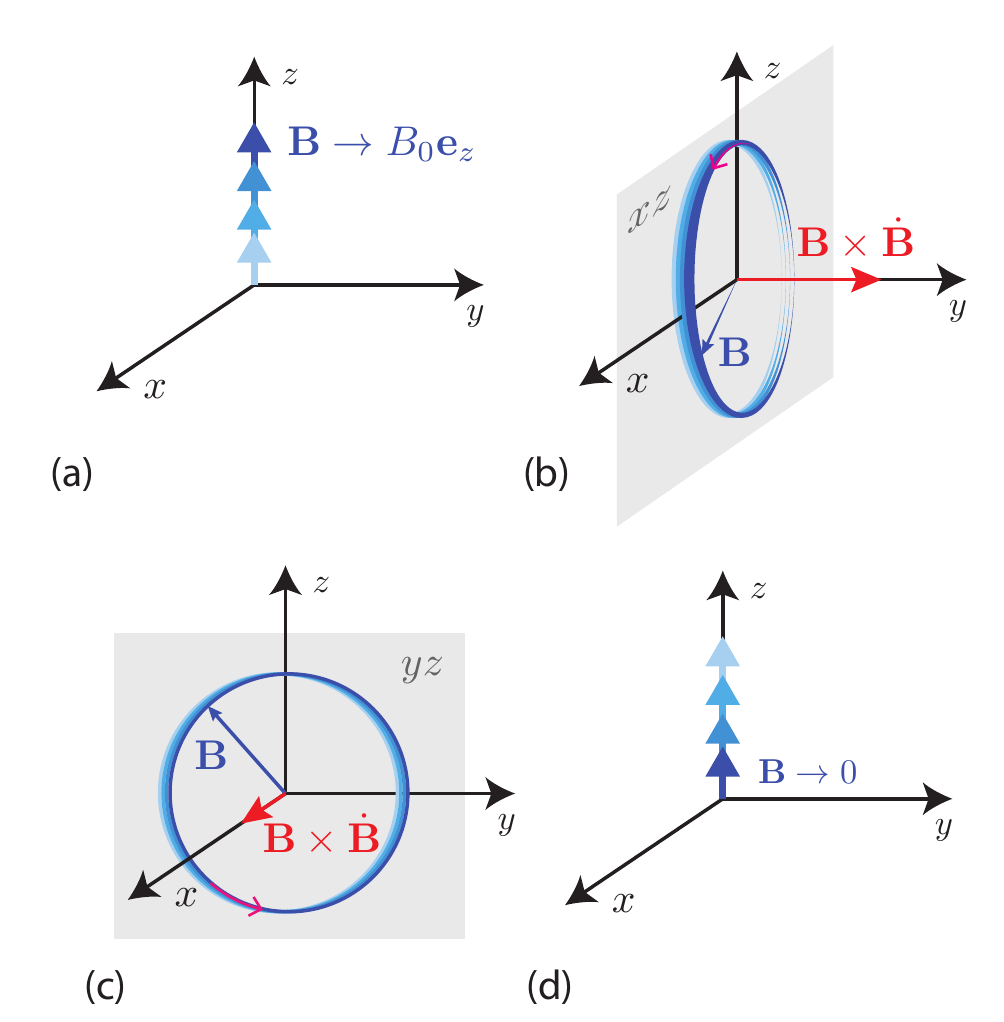} \tabularnewline
\end{tabular}
\par\end{centering}
\caption{\label{fig:Temporal-sequence}An example of a spin driven by an oscillating
magnetic field. The scheme involves four stages which include raising
of the magnetic field along the $z$ axis (a), subsequent rotation of
the magnetic field along the $y$ and $x$ axes (b), (c), and switching
off the magnetic field which again points along the $z$ axis (d). }
\end{figure}

The effect of the geometric phases can be measured using, for instance,
the following sequence illustrated in Fig.~\ref{fig:Temporal-sequence}.
Initially at $t=t_{0}$ the oscillating magnetic field and its derivative
are zero: $B\left(t_{0}\right)\rightarrow0$ and $\dot{B}\left(t_{0}\right)\rightarrow0$.
Therefore, according to the Eq.~(\ref{eq:micro_1}), the micromotion is absent at the initial time: $U_{\textrm{Micro}}\left(\omega t_{0}+\theta,t_{0}\right)=1$.
Subsequently, for $t_{0}<t<t_{1}$ the magnetic field strength increases
until reaching a steady-state value $B=B_{0}$. In doing so, the direction
of the magnetic field is kept fixed along the $z$ axis ($\mathbf{B}\left(t\right)\parallel\mathbf{e}_{z}$),
as indicated in Fig.~\ref{fig:Temporal-sequence} (a). The effective
Hamiltonian $H_{\textrm{eff}}$ is then zero. Therefore, apart from
the micromotion there is no other dynamics at this stage: $U_{\textrm{eff}}\left(t_{1},t_{0}\right)=1$.
In the next interval $t_{1}<t<t_{2}$ the magnetic field maintains
a constant amplitude $B_{0}$ and changes its direction, so non-zero
effective Hamiltonian $H_{\textrm{eff}}$ contributes to the temporal
evolution of the system. During that stage, the magnetic field may
perform a number rotations along different axes $\mathbf{n}_{j}$,
described by non-commuting unitary operators $U_{\textrm{eff}}\left(\mathbf{n}_{j},\varphi_{j}\right)$.
This is illustrated in Figs~\ref{fig:Temporal-sequence} (b)
and (c) showing two rotations: along the $y$ and $x$ axes ($\mathbf{n}_{j}=\mathbf{e}_{y},\,\mathbf{e}_{x}$).
In the final interval $t_{2}<t<t_{3}$, the magnetic field is decreasing
without changing its direction, so that $H_{\textrm{eff}}=0$ and
hence $U_{\textrm{eff}}\left(t_{3},t_{2}\right)=1$. At the final
time $t=t_{3}$, the magnetic field and its derivative go to zero
$B\left(t_{3}\right)\rightarrow0$ and $\dot{B}\left(t_{3}\right)\rightarrow0$
(see Fig.~\ref{fig:Temporal-sequence} (d)), so the micromotion
vanishes.

Since there is no micromotion at the initial and final times: $U_{\textrm{Micro}}\left(\omega t_{0}+\theta,t_{0}\right)=U_{\textrm{Micro}}\left(\omega t_{3}+\theta,t_{3}\right)=1$,
according to Eq.~(\ref{eq:phi-t-solution}) the state vector at the
final time is related to that at the initial time by a $\theta$-independent
effective evolution operator $U_{\textrm{eff}}\left(t_{3},t_{0}\right)=U_{\textrm{eff}}\left(t_{2},t_{1}\right)$:

\begin{equation}
\left|\phi_{\theta}\left(t_{3}\right)\right\rangle =U_{\textrm{eff}}\left(t_{3},t_{0}\right)\left|\alpha\right\rangle =U_{\textrm{eff}}\left(t_{2},t_{1}\right)\left|\alpha\right\rangle \label{eq:phi_solution-spin}
\end{equation}
In this way, the long-time dynamics of state vector is described by
the same effective evolution operator $U_{\textrm{eff}}\left(t_{3},t_{0}\right)$
for an arbitrary phase $\theta$ entering the Hamiltonian $H\left(\omega t+\theta,t\right)$.
This makes the scheme insensitive to the phase $\theta$ and a way the magnetic field is switched on and off.

It is noteworthy that the dynamics of a spin adiabatically following
a slowly changing magnetic field was considered by Berry~\cite{Berry:1984}.
In that case, an adiabatic elimination of the second spin component
provided a geometric (Berry) phase after a cyclic evolution. Such
a geometric phase is Abelian, because the effective dynamics involves
a single-spin component adiabatically following the magnetic field.

In the present situation relying on a fast oscillating magnetic field
with a changing direction, the spin is no longer adiabatically following
the magnetic field. Therefore, the spin degree of freedom is no longer
frozen and the emergence of the non-Abelian phases is possible.
The non-Abelian geometric phases arise now due to adiabatic elimination
of the extended space Floquet bands with $m\ne0$ (shown in 
Fig.~\ref{fig:Floquet-levels} and in Fig.~\ref{fig:spin-1/2}(a)
in Appendix \ref{sec:Spin--Relation to rotation frequency shift}),
rather than of the physical states, as it is usually the case~\cite{Wilczek:1984,Moody1986,Zee1988}.

\begin{figure}
\begin{centering}
\begin{tabular}{c}
\includegraphics[width=0.99\columnwidth]{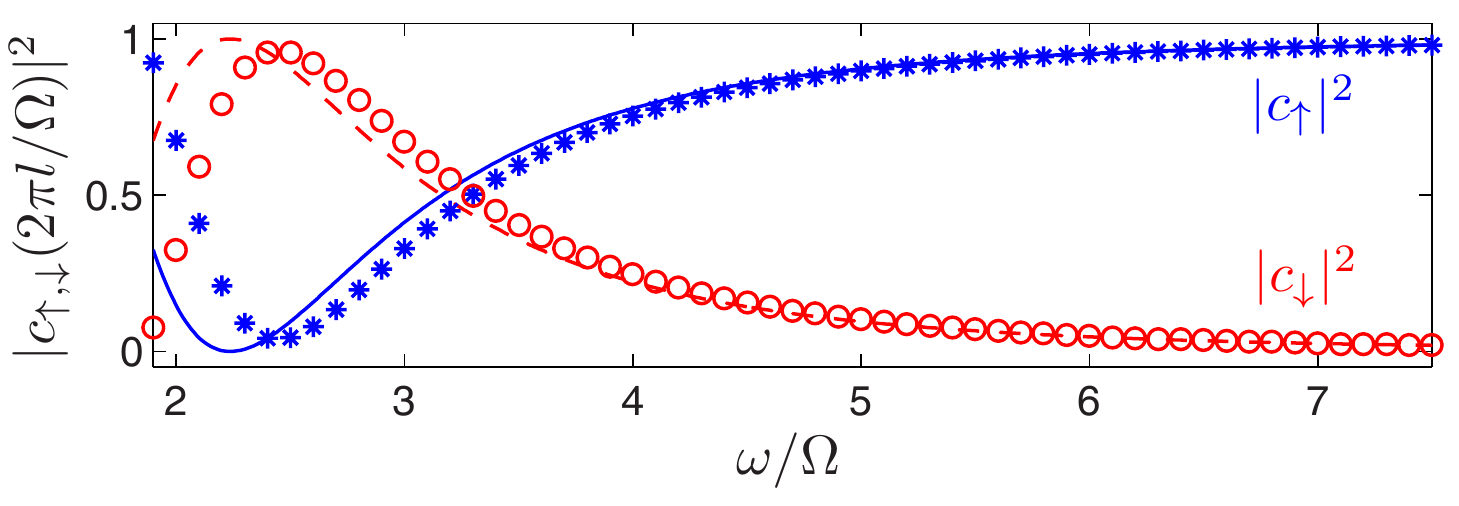} \tabularnewline
\end{tabular}
\par\end{centering}
\caption{\label{fig:spin_numeric} The probabilities $|c_{\uparrow}(2\pi l/\Omega)|^2$ and $|c_{\downarrow}(2\pi l/\Omega)|^2$ to find the spin in the states $\left|\uparrow\right\rangle$ (blue asterisk and solid line) and $\left|\downarrow\right\rangle$ (red circles and dashed line) after $l=10$ rotations of the magnetic field. The probabilities have been calculated numerically (symbols) and from the effective evolution Eqs.~(\ref{eq:fin_prob}) (continuous lines). The characteristic amplitude of the magnetic field is taken to be such that $g_F B/\Omega=1$.}
\end{figure}

Finally, let us compare the analytical expression (\ref{U_eff-spin-2}) for the effective dynamics with numerical simulations. For this we numerically calculate the exact evolution of a spin-$1/2$ particle governed by the Hamiltonian (\ref{eq:H-spin}), in which the amplitude of the oscillating magnetic field  $\mathbf{B}(t)=B\left[\mathbf{e}_z \cos(\Omega t)-\mathbf{e}_y \sin(\Omega t)\right]$ rotates in the $yz$ plane. After preparing the system in a spin-up state, $\left|\alpha\right\rangle= \left|\uparrow\right\rangle$, we allow the magnetic field vector $\mathbf{B}(t)$ to make $l=10$ rotations in the $yz$ plane. This transforms the state vector to a superposition of the spin-up and -down states: $c_{\uparrow}(t_l)\left|\uparrow\right\rangle+c_{\downarrow}(t_l)\left|\downarrow\right\rangle$, with $t_l=2\pi l/\Omega$. The corresponding probabilities $|c_{\uparrow}(2\pi l/\Omega)|^2$ and $|c_{\downarrow}(2\pi l/\Omega)|^2$ calculated numerically for  $l=10$ are depicted by asterisks and circles in Fig.~\ref{fig:spin_numeric}.  To remove the fast oscillations due to the micromotion operators, the numerical simulations have been performed by taking the values of the driving frequency $\omega$ such that $l\omega/\Omega$ remains integer.  

On the other hand, the effective evolution described by  Eq.~(\ref{U_eff-spin-2}) yields the following analytical expressions for these probabilities: 
\begin{subequations} \label{eq:fin_prob} 
\begin{align}
|c_{\uparrow}(2\pi l/\Omega)|^2  & = \cos^2 \left(\frac{\pi l g_F^2 B^2}{4\omega^2}\right) ,\label{eq:fin_prob_1}\\
|c_{\downarrow}(2\pi l/\Omega)|^2  & =\sin^2 \left(\frac{\pi l g_F^2 B^2}{4\omega^2}\right) .\label{eq:fin_prob_2}
\end{align}
\end{subequations}
As one can see from Fig.~\ref{fig:spin_numeric}, the numerical results agree well with the analytical ones (shown in solid lines) when the driving frequency $\omega$  exceeds considerably the frequency of the magnetic field rotation: $\omega \gg \Omega$.

\section{Concluding remarks \label{Conclusions}} 

We have considered a quantum system described by the Hamiltonian $H\left(\omega t+\theta,t\right)$
which is $2\pi$-periodic with respect to the first argument $\omega t+\theta$
and allows for an additional (slow) temporal dependence represented
by the second argument. The periodic time-dependence of the Hamiltonian
has been eliminated applying the extended-space formulation of the
Floquet theory~\cite{Guerin2003}. Consequently the original Schr\"{o}dinger-type
equation~(\ref{eq:Shroedinger-initial}) has been transformed into
an equivalent Schr\"{o}dinger-like equation of motion~(\ref{eq:evolution-1})
governed by the extended-space Hamiltonian~(\ref{eq:K-expansion})
containing \emph{only} a slow temporal dependence.

Using such an approach, Eq.~(\ref{eq:phi-t-solution}) has been obtained
describing the evolution of the system in terms of a long term dynamics
governed by the $\theta$-independent unitary operator $U_{\textrm{eff}}\left(t,t_{0}\right)$, 
as well as the $\theta$-dependent micromotion operators $U_{\textrm{Micro}}\left(\omega t+\theta,t\right)$
taken at the initial and final times, $t=t_{0}$ and $t$. The former
operator $U_{\textrm{eff}}\left(t,t_{0}\right)$ is determined by
the effective Hamiltonian $H_{\textrm{eff}}=H_{\textrm{eff}}\left(t\right)$
slowly changing in time. The latter $U_{\textrm{Micro}}\left(\omega t+\theta,t\right)$
not only describes the fast periodic motion, but also exhibits an
additional slow temporal dependence. 

We have provided a general framework for a combined analysis
of a high-frequency perturbation and slow changes in the periodic
driving. The micromotion operators and the effective Hamiltonian have
been systematically constructed in terms of a series in the powers
of $\omega^{-1}$. Analytical expressions~(\ref{eq:H_eff_expansion}), (\ref{eq:H_eff}),
and~(\ref{eq:U_kick_expansion}) give the expansions to the second
order in $\omega^{-1}$ inclusively.

In the limit of a strictly time-periodic Hamiltonian $H=H\left(\omega t+\theta\right)$,
the expansions reproduce the ones presented in previous studies~\cite{Eckardt2015,Goldman2014,Bukov2015,Mikami16PRB,Eckardt16-Review}.
Yet, in a more general situation considered here, the effective Hamiltonian
and the micromotion operators incorporate the dependence on the slow
time. Thus, they change their form during the course of the evolution.
Furthermore, the effective Hamiltonian and micromotion operators contain
additional second-order contributions emerging entirely from the slow
temporal dependence of the Fourier components composing the original
Hamiltonian in Eq.~(\ref{eq:H-periodic-expansion}). 

To show the effect of the additional terms on the dynamics, in Sec.~\ref{sec:Oscillating spin}
we have studied a spin $\hat{\mathbf{F}}$ in a magnetic field oscillating
rapidly along a slowly changing direction. If the changes in the orientation
of the magnetic field are not restricted to a single plane, the effective
evolution of the spin provides non-Abelian geometric phases.

The general theory is applicable to other driven systems, such
as periodically modulated optical lattices with a time-dependent forcing
strength. Such a situation is relevant to cold-atom experiments~\cite{Eckardt2010,Aidelsburger:2011,Arimondo2012,Hauke:2012,Windpassinger2013RPP,Struck:2013,Aidelsburger:2013,Jotzu2014,Aidelsburger14NP,Kennedy15,Flaschner16}. Shaking of optical lattices is known to renormalize inter-site tunneling amplitudes
\cite{eckardt05,lignier07,Arimondo2012,Bukov2015,Eckardt16-Review}. 
In the case of a slowly varying driving, the tunneling amplitudes acquire a slow temporal dependence. 
In Appendix \ref{sec:1d-lattice} this is illustrated for a one-dimensional optical lattice affected by a slowly changing shaking.

If the periodic modulation is absent at the initial
time $t_{0}$ and is slowly switched on afterwards, the micromotion
operator $U_{\textrm{Micro}}^{\dagger}\left(\omega t_{0}+\theta,t_{0}\right)$
reduces to a unit operator in Eq.~(\ref{eq:phi-t-solution}). In
that case, the temporal evolution of the system is described by the
effective Hamiltonian $H_{\textrm{eff}}=H_{\textrm{eff}}\left(t\right)$
slowly changing in time, and fast oscillating micromotion operator
$U_{\textrm{Micro}}\left(\omega t+\theta,t\right)$ calculated only
at the final time. 
\begin{acknowledgments}
The authors are grateful to I.~Bloch, A.~Eckardt, D.~Efimov, N.~Goldman, H.~Pu, J.~Ruseckas, and I.~Spielman
for helpful discussions and comments. This work was supported by the
Research Council of Lithuania under the Grant No. APP-4/2016. 
\end{acknowledgments}

\clearpage{}

\appendix

\section{Unitarity of the operator $U_{\textrm{Micro}}\left(\omega t+\theta,t\right)$\label{sec:Appendix-A:-Unitarity U}}

Here, we will show that the micromotion operator $U_{\textrm{Micro}}\left(\omega t+\theta,t\right)$
is unitary in the physical space. By definition, the operator $\hat{D}\left(t\right)$
is unitary in the extended space: 
\[
\hat{D}^{\dagger}\hat{D}=\sum_{n}\left|\overline{n}\right\rangle \mathbf{1}_{\mathscr{H}}\left\langle \overline{n}\right|\,,
\]
where $\mathbf{1}_{\mathscr{H}}$ is a unit operator in the physical
Hilbert space $\mathscr{H}$. On the other hand, using~(\ref{eq:D-expansion})
for $\hat{D}\left(t\right)$ and the fact that $\hat{P}_{l}\hat{P}_{m}=\hat{P}_{m+l}$,
one finds 
\[
\hat{D}^{\dagger}\hat{D}=\sum_{m,n}D^{\left(m\right)\dagger}D^{\left(n\right)}\hat{P}_{-m}\hat{P}_{n}=\sum_{m,l}D^{\left(m\right)\dagger}D^{\left(l+m\right)}\hat{P}_{l}\,.
\]
Comparing the two equations, one obtains the following condition for
$D^{\left(m\right)}$: 
\begin{equation}
\sum_{m=-\infty}^{\infty}D^{\left(m\right)\dagger}D^{\left(m+l\right)}=\mathbf{1}_{\mathscr{H}}\delta_{l0}.
\end{equation}
Consequently, one finds that the micromotion operator given by Eq.~(\ref{eq:U_Kick})
is indeed unitary: 
\begin{equation}
U_{\textrm{Micro}}^{\dagger}U_{\textrm{Micro}}=\sum_{m,n}D^{(m)\dagger}\left(t\right)D^{(n)}\left(t\right)e^{i\left(n-m\right)\omega t}=\mathbf{1}_{\mathscr{H}}\,.\label{eq:Unitarity-of-U}
\end{equation}


\section{Expansion of the effective Hamiltonian in the powers of $1/\omega$
\label{sec:Appendix-B:-Expansion}}

\subsection{Basic initial equations}
\begin{widetext}
It is convenient to represent the abstract extended-space Hamiltonian~(\ref{eq:K-expansion})
as 
\begin{equation}
\hat{K}=\hbar\omega\hat{N}+\hat{K}'\,,\quad\mathrm{with}\quad\hat{K}'=\sum_{m=-\infty}^{+\infty}H^{\left(m\right)}\hat{P}_{m}\,,\,,\label{eq:K-Appendix-A}
\end{equation}
where $\hat{N}$ is the ``number'' operator in the Floquet basis
given by Eq.~(\ref{eq:N}), and $\hat{P}_{m}$ is defined by~(\ref{eq:P_l}).
Below we will use some properties of the operators $\hat{P}_{m}$,
namely,
\begin{equation}
\left[\hat{P}_{m},\hat{N}\right]=-m\hat{P}_{m},\quad\hat{P}_{m}\hat{P}_{n}=\hat{P}_{m+n}\quad\textrm{and}\quad\left[\hat{P}_{m},\hat{P}_{n}\right]=0.\label{eq:relation-P}
\end{equation}

We are looking for a unitary transformation 
\begin{equation}
\hat{D}\left(t\right)=e^{-i\hat{S}\left(t\right)}\label{eq:U_D-Appendix-A}
\end{equation}
which makes the Hamiltonian $\hat{K}$ given Eq.~(\ref{eq:K-Appendix-A})
diagonal in the abstract extended space:

\begin{equation}
\hat{D}^{\dagger}\hat{K}\hat{D}-i\hbar\hat{D}^{\dagger}\dot{\hat{D}}=e^{i\hat{S}}\hat{K}e^{-i\hat{S}}-i\hbar e^{i\hat{S}}\frac{d}{dt}e^{-i\hat{S}}=\hat{K}_{D}\,,\label{eq:Block-diag-Appendix-A}
\end{equation}
where 
\begin{equation}
\hat{K}_{D}\left(t\right)=\hbar\omega\hat{N}+H_{\textrm{eff}}\left(t\right)\hat{P}_{0}\,.\label{eq:K_D-Appendix-A}
\end{equation}
is block diagonal.

\subsection{High-frequency expansion}

We are interested in a situation where the spectrum of $H_{\textrm{eff}}$
is confined in an energy range much smaller than the separation between
the Floquet bands $\hbar\omega$. In that case, the effective Hamiltonian
$H_{\textrm{eff}}$ can be expanded in powers of the inverse driving
frequency $1/\omega$: 
\begin{equation}
H_{\textrm{eff}}=H_{\textrm{eff}\left(0\right)}+H_{\textrm{eff}\left(1\right)}+H_{\textrm{eff}\left(2\right)}+\ldots\,.\label{eq:K_Df-expansion-Appendix-A}
\end{equation}
where the $j$-th term $H_{\textrm{eff}\left(j\right)}$ is of the
order of $1/\omega^{j}$. As we shall see later on, the zero-order
term $H_{\textrm{eff}\left(0\right)}$ coincides with the contribution
due to the zero-frequency component $H^{\left(0\right)}$ of the physical
Hamiltonian.

The Hermitian operator $\hat{S}$ featured in the unitary transformation
$\hat{D}$, Eq.~(\ref{eq:U_D-Appendix-A}), can also be expanded
in the powers of $1/\omega$ as: 
\begin{equation}
\hat{S}=\hat{S}_{\left(1\right)}+\hat{S}_{\left(2\right)}+\ldots\,,\label{eq:S_D-expansion-Appendix-A}
\end{equation}
where the expansion does not contain the zero-order term, because
the unitary operator $\hat{D}=\exp\left(-i\hat{S}\right)$ should
approach the unity in a very high-frequency limit.

We are also looking for the $1/\omega$ power expansion of the unitary
operator $\hat{D}$: 
\begin{equation}
\hat{D}=\hat{D}_{\left(0\right)}+\hat{D}_{\left(1\right)}+\hat{D}_{\left(2\right)}+\ldots\,,\label{eqU_D-expansion-Appendix-A}
\end{equation}
with \begin{subequations} 
\begin{equation}
\hat{D}_{\left(0\right)}=\mathbf{1}_{\mathscr{L}}\,,\quad\hat{D}_{\left(1\right)}=-i\hat{S}_{\left(1\right)}\,,
\end{equation}
\begin{equation}
\hat{D}_{\left(2\right)}=-i\hat{S}_{\left(2\right)}-\frac{1}{2}\left[\hat{S}_{\left(1\right)}\right]^{2}\,.
\end{equation}
\end{subequations}

\subsection{Determination of the high-frequency expansion of $H_{\textrm{eff}}$
and $\hat{S}$}

To find the high-frequency expansion of $H_{\textrm{eff}}$ and $\hat{S}$,
let us express $e^{i\hat{S}}\hat{K}e^{-i\hat{S}}$ in the powers of
$\hat{S}$ as 
\begin{equation}
e^{i\hat{S}}\hat{K}e^{-i\hat{S}}=\hat{K}+i\left[\hat{S},\hat{K}\right]-\frac{1}{2!}\left[\hat{S},\left[\hat{S},\hat{K}\right]\right]-\frac{i}{3!}\left[\hat{S},\left[\hat{S},\left[\hat{S},\hat{K}\right]\right]\right]+\ldots\,,\label{eq:K_D-expansion-Appendix-A}
\end{equation}
Also let us calculate time derivative caused by the operator $\hat{D}$
time-dependence. We restrict ourselves up to third-order terms: 
\begin{equation}
-i\hbar\hat{D}^{\dagger}\dot{\hat{D}}=-\hbar\dot{\hat{S}}_{\left(1\right)}-\hbar\dot{\hat{S}}_{\left(2\right)}-\frac{i\hbar}{2}\left[\hat{S}_{\left(1\right)},\dot{\hat{S}}_{\left(1\right)}\right]+\mathcal{O}\left(\omega^{-3}\right).
\end{equation}
Using~(\ref{eq:K_D-Appendix-A}) and~(\ref{eq:K-Appendix-A}), sum
of the above equations reads as
\begin{equation}
\begin{aligned}H_{\textrm{eff}}\hat{P}_{0}= & \underbrace{\hat{K}'+i\left[\hat{S}_{\left(1\right)},\hbar\omega\hat{N}\right]}_{\textrm{zero order terms}}+\underbrace{i\left[\hat{S}_{\left(1\right)},\hat{K}'\right]+i\left[\hat{S}_{\left(2\right)},\hbar\omega\hat{N}\right]-\frac{1}{2!}\left[\hat{S}_{\left(1\right)},\left[\hat{S}_{\left(1\right)},\hbar\omega\hat{N}\right]\right]-\hbar\dot{\hat{S}}_{\left(1\right)}}_{\textrm{first order terms}}\\
 & \underbrace{+i\left[\hat{S}_{\left(2\right)},\hat{K}'\right]-\frac{1}{2!}\left[\hat{S}_{\left(1\right)},\left[\hat{S}_{\left(2\right)},\hbar\omega\hat{N}\right]\right]-\frac{1}{2!}\left[\hat{S}_{\left(2\right)},\left[\hat{S}_{\left(1\right)},\hbar\omega\hat{N}\right]\right]-\frac{1}{2!}\left[\hat{S}_{\left(1\right)},\left[\hat{S}_{\left(1\right)},\hat{K}'\right]\right]}_{\textrm{second order terms}}\\
 & \underbrace{-\frac{i}{3!}\left[\hat{S}_{\left(1\right)},\left[\hat{S}_{\left(1\right)},\left[\hat{S}_{\left(1\right)},\hbar\omega\hat{N}\right]\right]\right]+i\left[\hat{S}_{\left(3\right)},\hbar\omega\hat{N}\right]-\hbar\dot{\hat{S}}_{\left(2\right)}-\frac{i\hbar}{2}\left[\hat{S}_{\left(1\right)},\dot{\hat{S}}_{\left(1\right)}\right]}_{\textrm{second order terms}}\\
 & +\mathcal{O}\left(\omega^{-3}\right)\,.
\end{aligned}
\label{eq:K_D-expansion-alt-Appendix-A}
\end{equation}

Since $\hat{D}$ has a block-diagonal form~(\ref{eq:D-expansion}),
the Hermitian operator $\hat{S}$ should have the same form: 
\begin{equation}
\hat{S}=\sum_{m=-\infty}^{+\infty}S^{\left(m\right)}\hat{P}_{m}\,.\label{eq:S_D-expansion-P-Appendix-A}
\end{equation}

\subsection{Zero order for $H_{\textrm{eff}}$ }

In the lowest order in $1/\omega$ one finds 
\begin{equation}
H_{\textrm{eff}\left(0\right)}\hat{P}_{0}=\hat{K}'+i\hbar\omega\left[\hat{S}_{\left(1\right)},\hat{N}\right]\,,\label{eq:K_D-expansion-0-order-Appendix-A}
\end{equation}
Expanding $\hat{K}'$ and $\hat{S}_{\left(1\right)}$ in terms of
the shift operators $\hat{P}_{m}$, the above equation yields 
\begin{equation}
H_{\textrm{eff}\left(0\right)}\hat{P}_{0}=\sum_{m=-\infty}^{+\infty}\left(H^{\left(m\right)}-im\hbar\omega S_{\left(1\right)}^{\left(m\right)}\right)\hat{P}_{m}\,.\label{eq:K_D-expansion-0-order-1-Appendix-A}
\end{equation}
Thus the zero-order Hamiltonian reads as
\begin{equation}
H_{\textrm{eff}\left(0\right)}=H^{\left(0\right)}.\label{eq:H_eff_0-Appendix-A}
\end{equation}

On the other hand, Eq.~(\ref{eq:K_D-expansion-0-order-Appendix-A})
provides the following result for the first-order contribution to
the Hermitian transformation exponent $\hat{S}$: 
\begin{equation}
\hat{S}_{\left(1\right)}=\frac{1}{i\hbar\omega}\sum_{m\ne0}\frac{1}{m}H^{\left(m\right)}\hat{P}_{m}\equiv\frac{1}{i\hbar\omega}\sum_{m=1}^{+\infty}\frac{1}{m}\left(H^{\left(m\right)}\hat{P}_{m}-H^{\left(-m\right)}\hat{P}_{-m}\right)\,.\label{eq:S_D1-Appendix-A}
\end{equation}
This is consistent with the first-order terms presented in Appendix
C of Ref.~\cite{Goldman2014}. Note that Eq.~(\ref{eq:K_D-expansion-0-order-Appendix-A})
does not define $S_{\left(1\right)}^{\left(0\right)}$, so we have
taken $S_{\left(1\right)}^{\left(0\right)}=0$. More generally, in
what follows we shall assume that $S_{\left(n\right)}^{\left(0\right)}=0$
in all orders $n$. In the following, we shall see that this assumption
is consistent also in higher orders of perturbation. Additionally,
we get the first-order term for the expansion of the unitary operator:
\begin{equation}
\hat{D}_{\left(1\right)}=\frac{-1}{\hbar\omega}\sum_{m\ne0}\frac{1}{m}H^{\left(m\right)}\hat{P}_{m}\,.\label{eq:U_D1}
\end{equation}

\subsection{First order for $H_{\textrm{eff}}$}

In the next order in $1/\omega$ one has 
\begin{equation}
H_{\textrm{eff}\left(1\right)}\hat{P}_{0}=i\left[\hat{S}_{\left(1\right)},\hat{K}'\right]+i\left[\hat{S}_{\left(2\right)},\hbar\omega\hat{N}\right]-\frac{1}{2!}\left[\hat{S}_{\left(1\right)},\left[\hat{S}_{\left(1\right)},\hbar\omega\hat{N}\right]\right]-\hbar\dot{\hat{S}}_{\left(1\right)}\,.\label{eq:expansion-next-order-Appendix-A}
\end{equation}
Combining Eqs.~(\ref{eq:H_eff_0-Appendix-A})-(\ref{eq:S_D1-Appendix-A})
for $H_{\textrm{eff}\left(0\right)}$ and $\hat{S}_{\left(1\right)}$
with auxiliary relationships~(\ref{eq:relation-P}), the above equation
simplifies to 
\begin{equation}
\begin{aligned}H_{\textrm{eff}\left(1\right)}\hat{P}_{0}= & \frac{1}{2\hbar\omega}\sum_{m \neq 0}\sum_{n \neq 0}\frac{\left[H^{\left(m\right)},H^{\left(n\right)}\right]}{m}\hat{P}_{m+n}+\frac{1}{\hbar\omega}\sum_{m\neq0}\frac{\left[H^{\left(m\right)},H^{\left(0\right)}\right]}{m}\hat{P}_{m}\\
 & -i\hbar\omega\sum_{m\neq0}mS_{\left(2\right)}^{\left(m\right)}\hat{P}_{m}-\frac{1}{i\omega}\sum_{m\ne0}\frac{1}{m}\dot{H}^{\left(m\right)}\hat{P}_{m}
\end{aligned}
\,.
\end{equation}
Thus first-order effective Hamiltonian is given by 
\begin{equation}
H_{\textrm{eff}\left(1\right)}=\frac{1}{\hbar\omega}\sum_{m=1}^{+\infty}\frac{1}{m}\left[H^{\left(m\right)},H^{\left(-m\right)}\right]\,.
\end{equation}
On the other hand, the second order of the transformation exponent
operator reads 
\begin{equation}
\hat{S}_{\left(2\right)}=\sum_{m\neq0}S_{\left(2\right)}^{\left(m\right)}\hat{P}_{m}\label{eq:S_2-App}
\end{equation}
where 
\begin{equation}
S_{\left(2\right)}^{\left(m\right)}=\frac{1}{2im\left(\hbar\omega\right)^{2}}\left\{ \frac{1}{m}\left[H^{\left(m\right)},H^{\left(0\right)}\right]+\sum_{n\neq0}\frac{1}{n}\left[H^{\left(n\right)},H^{\left(m-n\right)}\right]\right\} +\frac{\hbar}{\left(\hbar\omega\right)^{2}m^{2}}\dot{H}^{\left(m\right)}\,.
\end{equation}
The second order term of the unitary operator takes the form 
\begin{equation}
\begin{aligned} & \hat{D}_{\left(2\right)}=\frac{1}{2\left(\hbar\omega\right)^{2}}\left[\sum_{m\neq0}\sum_{n\neq0}\hat{P}_{m+n}\frac{H^{\left(m\right)}H^{\left(n\right)}}{nm}-\sum_{m\ne0}\hat{P}_{m}\right.\\
 & \left.\times\left\{ \frac{\left[H^{\left(m\right)},H^{\left(0\right)}\right]+2i\hbar\dot{H}^{\left(m\right)}}{m^{2}}+\sum_{n\neq0}\frac{\left[H^{\left(n\right)},H^{\left(m-n\right)}\right]}{mn}\right\} \right]
\end{aligned}
\,.\label{eq:U_D2}
\end{equation}

\subsection{Second order for $H_{\textrm{eff}}$}

In the next order in $1/\omega$ one has 
\begin{equation}
\begin{aligned} & H_{\textrm{eff}\left(2\right)}\hat{P}_{0}=i\left[\hat{S}_{\left(2\right)},\hat{K}'\right]+i\left[\hat{S}_{\left(3\right)},\hbar\omega\hat{N}\right]-\frac{1}{2}\left[\hat{S}_{\left(1\right)},\left[\hat{S}_{\left(1\right)},\hat{K}'\right]\right]-\frac{1}{2}\left[\hat{S}_{\left(1\right)},\left[\hat{S}_{\left(2\right)},\hbar\omega\hat{N}\right]\right]\\
 & -\frac{1}{2}\left[\hat{S}_{\left(2\right)},\left[\hat{S}_{\left(1\right)},\hbar\omega\hat{N}\right]\right]-\frac{i}{6}\left[\hat{S}_{\left(1\right)},\left[\hat{S}_{\left(1\right)},\left[\hat{S}_{\left(1\right)},\hbar\omega\hat{N}\right]\right]\right]-\hbar\dot{\hat{S}}_{\left(2\right)}-\frac{i\hbar}{2}\left[\hat{S}_{\left(1\right)},\dot{\hat{S}}_{\left(1\right)}\right]\,.
\end{aligned}
\label{eq:App_H_eff_2}
\end{equation}
Each term in the right-hand-side of the Eq.~\eqref{eq:App_H_eff_2} can be
considered as a sum $\sum_{m}X^{\left(m\right)}\hat{P}_{m}$. To find
$H_{\textrm{eff}\left(2\right)}$ we need to determine only the operator
$X^{\left(0\right)}$. Hence, the third term in the right-hand-side of the Eq.~\eqref{eq:App_H_eff_2}
gives 
\begin{equation}
-\frac{1}{2}\left[\hat{S}_{\left(1\right)},\left[\hat{S}_{\left(1\right)},\hat{K}'\right]\right]_{\hat{P}_{0}}=\frac{1}{2\left(\hbar\omega\right)^{2}}\sum_{m\neq0}\sum_{n\neq0}\frac{1}{mn}\left[\left[H^{\left(n\right)},H^{\left(m-n\right)}\right],H^{\left(-m\right)}\right]\,.\label{eq:Sd2_projection_to_p0}
\end{equation}
The second and seventh terms on the r.h.s. of the Eq.~\eqref{eq:App_H_eff_2}
give a zero contribution. The first, fourth and fifth terms together
also do not contribute. The sixth and eighth terms give 
\begin{equation}
-\frac{i}{6}\left[\hat{S}_{\left(1\right)},\left[\hat{S}_{\left(1\right)},\left[\hat{S}_{\left(1\right)},\hbar\omega\hat{N}\right]\right]\right]_{\hat{P}_{0}}=\frac{-1}{2\left(\hbar\omega\right)^{2}}\sum_{m\neq0}\sum_{n\neq\{0,m\}}\frac{\left[\left[H^{\left(n\right)},H^{\left(m-n\right)}\right],H^{\left(-m\right)}\right]}{3mn}\,,\label{eq:Sd2_projection_to_p0-1}
\end{equation}
\begin{equation}
-\frac{i\hbar}{2}\left[\hat{S}_{\left(1\right)},\dot{\hat{S}}_{\left(1\right)}\right]_{\hat{P}_{0}}=-\frac{i\hbar}{2\left(\hbar\omega\right)^{2}}\sum_{m\neq0}\frac{1}{m^{2}}\left[H^{\left(-m\right)},\dot{H}^{\left(m\right)}\right]\,.
\end{equation}
In this way, the second order of the effective Hamiltonian reads

\begin{equation}
H_{\textrm{eff}\left(2\right)}=\sum_{m\neq0}\left\{ \frac{\left[H^{\left(-m\right)},\left[H^{\left(0\right)},H^{\left(m\right)}\right]\right]-i\hbar\left[H^{\left(-m\right)},\dot{H}^{\left(m\right)}\right]}{2\left(m\hbar\omega\right)^{2}}+\sum_{n\neq\{0,m\}}\frac{\left[H^{\left(-m\right)},\left[H^{\left(m-n\right)},H^{\left(n\right)}\right]\right]}{3mn\left(\hbar\omega\right)^{2}}\right\} \,.\label{eq:H_eff_2-1}
\end{equation}

\subsection{Power expansion of the operator $U_{\textrm{Micro}}\left(\theta^{\prime},t\right)$}

The time dependence of the operator $U_{\textrm{Micro}}\left(\theta^{\prime},t\right)$
can be recovered from the expansion of the unitary operator $\hat{D}$:
\begin{equation}
\begin{aligned} & U_{\textrm{Micro}}\left(\theta^{\prime},t\right)\approx\mathbf{1}-\frac{1}{\hbar\omega}\sum_{m\ne0}\frac{1}{m}H^{\left(m\right)}e^{im\theta^{\prime}}+\frac{1}{2\left(\hbar\omega\right)^{2}}\sum_{m\neq0}\sum_{n\neq0}e^{i\left(m+n\right)\theta^{\prime}}\frac{H^{\left(m\right)}H^{\left(n\right)}}{nm}\\
 & +\frac{1}{2\left(\hbar\omega\right)^{2}}\sum_{m\ne0}e^{im\theta^{\prime}}\left\{ \frac{\left[H^{\left(0\right)},H^{\left(m\right)}\right]-2i\hbar\dot{H}^{\left(m\right)}}{m^{2}}-\sum_{n\neq0}\frac{\left[H^{\left(n\right)},H^{\left(m-n\right)}\right]}{nm}\right\} \,.
\end{aligned}
\label{eq:U_kick}
\end{equation}

\subsection{Power expansion of the operator $S_{\textrm{Micro}}\left(t\right)$}

The expansion of the Hermitian operator $S_{\textrm{Micro}}=S_{\textrm{Micro}\left(1\right)}+S_{\textrm{Micro}\left(2\right)}+\mathcal{O}\left(\omega^{-3}\right)$
defined as the exponential form $U_{\textrm{Micro}}=\exp\left[-iS_{\textrm{Micro}}\right]$
can be recovered from the expansion of the operator $\hat{S}$ by
taking $\hat{P}_{m}\rightarrow\exp\left(im\theta^{\prime}\right)$:
\begin{equation}
\begin{aligned}S_{\textrm{Micro}\left(1\right)}\left(\theta^{\prime},t\right)= & \frac{1}{i\hbar\omega}\sum_{m\ne0}\frac{1}{m}H^{\left(m\right)}e^{im\theta^{\prime}}\\
S_{\textrm{Micro}\left(2\right)}\left(\theta^{\prime},t\right)= & \frac{1}{2i\left(\hbar\omega\right)^{2}}\sum_{m\ne0}\left\{ \frac{1}{m^{2}}\left[H^{\left(m\right)},H^{\left(0\right)}\right]+\sum_{n\neq0}\frac{1}{mn}\left[H^{\left(n\right)},H^{\left(m-n\right)}\right]+\frac{2i\hbar}{m^{2}}\dot{H}^{\left(m\right)}\right\} e^{im\theta^{\prime}}\,.
\end{aligned}
\label{eq:S_kick}
\end{equation}
\end{widetext}

\section{Floquet effective Hamiltonian of a one-dimensional optical lattice with a time-dependent driving amplitude \label{sec:1d-lattice}}

Let us consider the atomic dynamics in a one-dimensional shaken optical lattice with a changing driving. In the laboratory frame, the periodic potential   
\begin{equation}
\label{eq:1d_latt_pot V}
V_{\mathrm{lab}}(t)= \frac{V_0}{2} \cos\left(2k_L\left[ x- X_0\left(t\right)\right]\right)
\end{equation}
is characterized by a lattice constant $b=\pi/k_L$. A temporal dependence of the lattice displacement $X_0\left(t\right)$ depends on the shaking protocol.
Here we consider a situation where a pair of counter-propagating laser beams creating the optical lattice are obtained by splitting a laser beam 
into two. A small time-dependent frequency difference $\Delta\nu(t)$ between the two split beams is produced using 
an acousto-optic modulator. This results in the lattice moving with a velocity $v(t) =  b\Delta\nu(t) $~\cite{Arimondo2012,BenDahan1996,lignier07,Madison1997,Madison1998,Niu1996,Sias2008}, so that
 \begin{equation}
\label{eq:X_0}
X_0\left(t\right)=b \int_{t_0}^t \mathrm{d}\tau \Delta\nu(\tau). 
\end{equation}
To produce the shaken optical lattice, we take a quasi-periodic modulation of $\Delta\nu(t)$ characterized by the frequency $\omega$, phase $\theta$, and slowly changing amplitude $f(t)$:   
 \begin{equation}
\label{eq:nu(t)}
\Delta \nu(t)=f(t) \sin(\omega t+ \theta ).
\end{equation}

For a sufficiently deep lattice potential, $V_0\gg E_{\mathrm{rec}}$, the optical lattice can be described by the tight-binding model, where $E_{\mathrm{rec}}=\hbar^2k_L^2/2M$ is the recoil momentum and $M$ is the atomic mass. The tight-binding Hamiltonian of the driven optical lattice reads as in a co-moving frame \cite{Arimondo2012}
\begin{equation}
\label{eq:H_1d_lattice}
H_{\textrm{dr}}(t)=-J\sum_{l=-\infty}^{\infty}\left(a^{\dagger}_la_{l-1} + a^{\dagger}_l a_{l+1} \right)+\sum_{l=-\infty}^{\infty}V_l(t) a^{\dagger}_l a_l ,
\end{equation}
where
 $J$ is a tunneling matrix element, $a^{\dagger}_l$ ($a_l$) is a creation (annihilation) operator for an atom at a lattice site $l$. Here also
 \begin{equation}
\label{eq:1d_latt_pot}
V_l(t)= lb  M\ddot{X}_0\left(t\right)
\end{equation}
is a modulated onsite potential.

To eliminate the on-site potential $V_l(t)$ proportional to the driving frequency $\omega$, one can apply a unitary transformation 
 \begin{equation}
\label{eq:U_dr}
U_{\textrm{dr}}(t)=\exp \left(-i\hbar^{-1} \sum_l lb M \dot{X}_0\left(t \right) a^{\dagger}_l a_l\right)
\end{equation}
 to the original Hamiltonian (\ref{eq:H_1d_lattice}), giving:
\begin{equation}
\begin{aligned}
H(t) & =U_{\textrm{dr}}^{\dagger}H_{\textrm{dr}}U_{\textrm{dr}}-i\hbar U_{\textrm{dr}}^{\dagger} \dot{U}_{\textrm{dr}}\\
 & =  -J \sum_{l=-\infty}^{\infty} \left(e^{i \varphi(t)}a^{\dagger}_la_{l-1} +e^{-i \varphi(t)}a^{\dagger}_la_{l+1} \right),
\end{aligned}
\label{eq:Htr_1d_lattice}
\end{equation}
with  
\begin{equation}
\label{eq:theta}
\varphi(t)=-\hbar^{-1} b M \dot{X}_0 = -\hbar^{-1}b^2 M f(t) \sin(\omega t+\theta).
\end{equation}

The transformed Hamiltonian (\ref{eq:Htr_1d_lattice}) no longer contains the large driving amplitude proportional to $\ddot{X}_0\left(t \right) \propto \omega$,  making its Fourier components $H^{(m)}(t)$ independent on the expansion parameter $\omega^{-1}$. The driving force is now captured by the time-dependent Peierls phase $\varphi(t)$.

Employing the relation
\begin{equation}
\label{eq:bessel}
\exp(i r \sin(s))= \sum_{m=-\infty}^{\infty} \mathcal{J}_m(r)\exp(ims),
\end{equation}
one obtains the Fourier components of the transformed Hamiltonian (\ref{eq:Htr_1d_lattice}):
\begin{equation}
\label{eq:1d_latt_fourier}
\begin{aligned}
H^{(m)}(t) & = -J \mathcal{J}_m \left(\frac{b^2 M f(t)}{\hbar} \right) \\
& \times \sum_{l=-\infty}^{\infty} \left((-1)^m a^{\dagger}_l a_{l-1} + a^{\dagger}_la_{l+1} \right),
\end{aligned}
\end{equation}
where $\mathcal{J}_m$ denotes a Bessel function of an integer order $m$.

According to Eqs.~(\ref{eq:H_eff_0}) and (\ref{eq:1d_latt_fourier}), the zero-order effective Hamiltonian has a form of the original Hamiltonian for the undriven system:   
\begin{equation}
\label{eq:latt_H0}
H_{\textrm{eff}(0)}= -J^{\prime}(t) \sum_{l=-\infty}^{\infty}\left(a^{\dagger}_la_{l-1} + a^{\dagger}_l a_{l+1} \right)\,,
\end{equation}
where the tunneling matrix element 
\begin{equation}
\label{eq:J^prime}
J^{\prime}(t)= J \mathcal{J}_0 \left(\frac{b^2 M f(t)}{\hbar} \right) 
\end{equation}
is rescaled by the Bessel function.
Unlike in the previous studies \cite{eckardt05,lignier07,Arimondo2012,Bukov2015,Eckardt16-Review}, the emerging Bessel function $ \mathcal{J}_0 \left(\frac{b^2 M f(t)}{\hbar} \right) $ changes in time due to the slow time-dependence of the driving. Note that the first and second order corrections for the effective Hamiltonian (\ref{eq:H_eff_1}), (\ref{eq:H_eff_2}) are zero:  $H_{\textrm{eff}(1)}=H_{\textrm{eff}(2)}=0$.

\section{Spin in an oscillating magnetic field: Relation to the rotation frequency
shift\label{sec:Spin--Relation to rotation frequency shift}}

In Sec.~\ref{sec:Oscillating spin} we have considered the
spin in the fast oscillating magnetic field $\mathbf{B}\left(t\right)\cos\left(\omega t+\theta\right)$
with the slowly varying amplitude $\mathbf{B}\left(t\right)$. Here,
we will show that the acquired geometric phases stem from the rotational
frequency shift~\cite{birula97rfs} representing a correction to
it. For this let us consider a case where the oscillating magnetic
field rotates at a constant angular frequency $\Omega$ around the
$z$ axis: $\mathbf{B}\left(t\right)=B\left[\cos\left(\Omega t\right)\mathbf{e}_{x}+\sin\left(\Omega t\right)\mathbf{e}_{y}\right]$.
According to Eq.~(\ref{eq:spin_effect}), a non-zero contribution
to the effective Hamiltonian emerges due to the rotation of $\mathbf{B}\left(t\right)$,
giving 
\begin{equation}
H_{\textrm{eff}}=H_{\textrm{eff}\left(2\right)}^{\left(2\right)}=\frac{g_{F}^{2}B^{2}\Omega}{4\omega^{2}}F_{3}.\label{eq:H_eff-xy}
\end{equation}

\begin{figure}
\begin{centering}
\begin{tabular}{c}
\includegraphics[width=0.99\columnwidth]{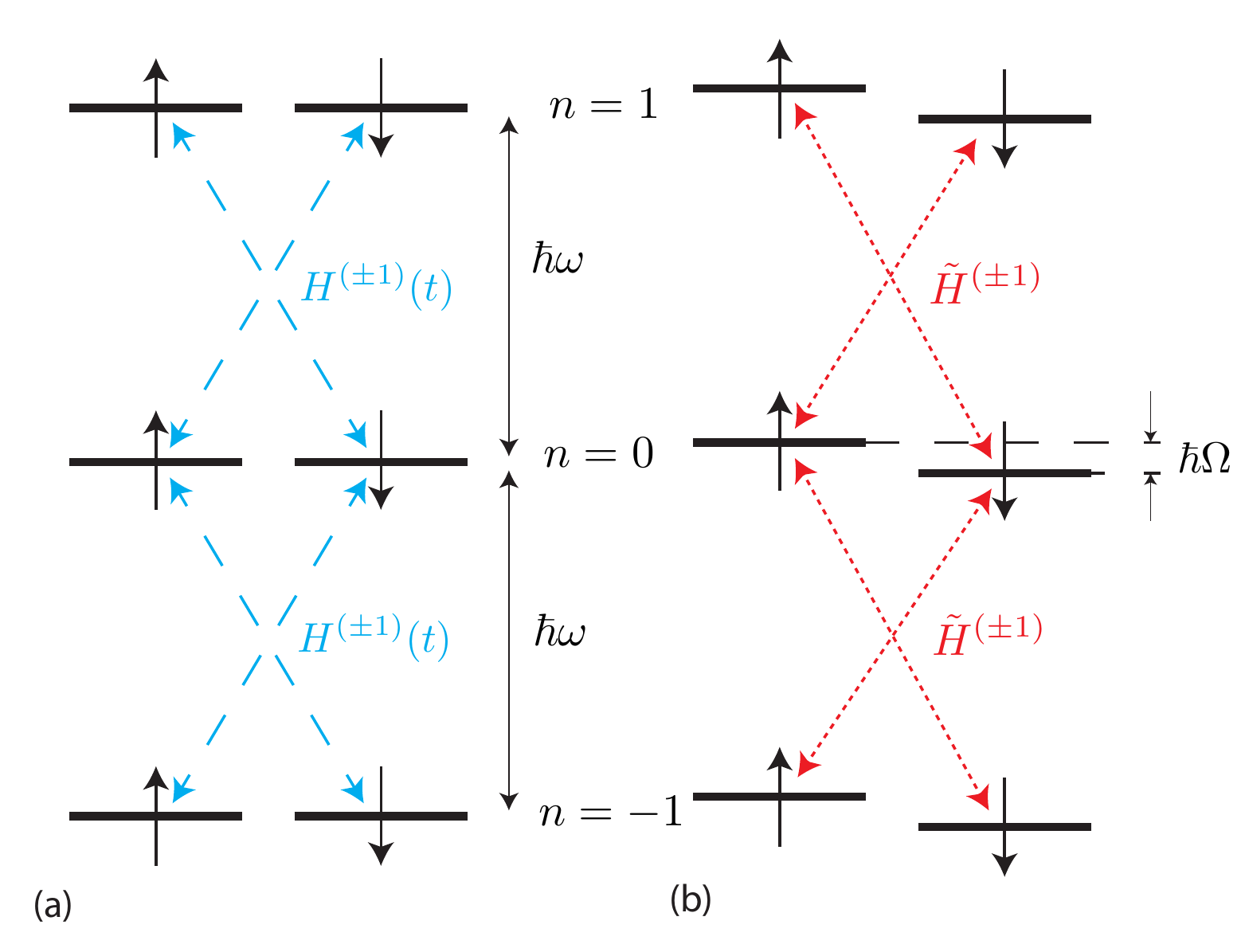} \tabularnewline
\end{tabular}
\par\end{centering}
\caption{\label{fig:spin-1/2}A Floquet picture for a spin $1/2$ system in
an oscillating magnetic field with a slowly rotating direction $\mathbf{B}\left(t\right)=B\left[\cos\left(\Omega t\right)\mathbf{e}_{x}+\sin\left(\Omega t\right)\mathbf{e}_{y}\right]$
in the laboratory (a) and rotating (b) frames. Dotted (red) arrows
represent quantum jumps between different Floquet bands in the rotating
frame described by the operators $\tilde{H}^{\left(\pm1\right)}=g_{F}BF_{1}/2$.
Dashed (blue) arrows represent the corresponding quantum jumps in
the laboratory frame described by the time-dependent operator $H^{\left(\pm1\right)}=g_{F}B\left[F_{1}\cos\left(\Omega t\right)+F_{2}\sin\left(\Omega t\right)\right]$
. The transition from the laboratory to the rotating frames introduces
a splitting of the spin-up and -down states by the amount $\hbar\Omega$,
as illustrated in (b). This leads to the second-order correction to
the Zeeman term given by Eq.~(\ref{eq:rot_eff_2}).}
\end{figure}

Alternatively, one can apply to the original Hamiltonian $H\left(\omega t+\theta,t\right)$
a unitary transformation $U_{\Omega t}=e^{-\frac{i}{\hbar}F_{3}\Omega t}$
rotating the spin along the $z$ axes by the angle $\Omega t$. The
transformed Hamiltonian then reads as
\begin{equation}
\begin{aligned}
\tilde{H}\left(\omega t+\theta,t\right) & =U_{\Omega t}^{\dagger}HU_{\Omega t}-i\hbar U_{\Omega t}^{\dagger}\dot{U}_{\Omega t}\\
 & =g_{F}F_{1}B\cos\left(\omega t+\theta\right)-F_{3}\Omega\,.
\end{aligned}
\label{eq:H_eff-xy-transformed}
\end{equation}
Therefore, in the new frame the oscillating magnetic field vector is
oriented in the $x$ direction and thus no longer rotates. Additionally,
a Zeeman term $-F_{3}\Omega$ appears due to the rotational frequency
shift~\cite{birula97rfs}. This is illustrated in Fig.~\ref{fig:spin-1/2}
for a spin-$1/2$ case. Non-zero Fourier components of the transformed
Hamiltonian are $\tilde{H}^{\left(0\right)}=-F_{3}\Omega$ and $\tilde{H}^{\left(-1\right)}=\tilde{H}^{\left(1\right)}=g_{F}F_{1}B/2$.
Therefore a high frequency expansion of the effective Hamiltonian
$\tilde{H}_{\textrm{eff}}$ reads in the rotating frame using Eqs.~(\ref{eq:H_eff})
\begin{subequations} \label{eq:rot_eff} 
\begin{align}
\tilde{H}_{\textrm{eff}\left(0\right)}= & -F_{3}\Omega,\label{eq:rot_eff_0}\\
\tilde{H}_{\textrm{eff}\left(1\right)}= & \mathbf{0}_{\mathscr{H}},\label{eq:rot_eff_1}\\
\tilde{H}_{\textrm{eff}\left(2\right)}= & \frac{1}{\left(\hbar\omega\right)^{2}}\left[\tilde{H}^{\left(1\right)},\left[\tilde{H}^{\left(0\right)},\tilde{H}^{\left(1\right)}\right]\right]=\frac{g_{F}^{2}B^{2}\Omega}{4\omega^{2}}F_{3}.\label{eq:rot_eff_2}
\end{align}
\end{subequations}The second order term $\tilde{H}_{\textrm{eff}\left(2\right)}$
in Eq.~(\ref{eq:rot_eff_2}) represents a correction to the rotational
energy shift. The term $\tilde{H}_{\textrm{eff}\left(2\right)}$ coincides
with Eq.~(\ref{eq:H_eff-xy}) for the shift in the laboratory frame
due to the changes in the direction of the oscillating magnetic field.
Yet in the rotating frame the effective Hamiltonian $\tilde{H}_{\textrm{eff}}$
also acquires the zero order term $\tilde{H}_{\textrm{eff}\left(0\right)}$
given by Eq.~(\ref{eq:rot_eff_0}), which is absent in the laboratory
frame. To eliminate $\tilde{H}_{\textrm{eff}\left(0\right)}$ one
needs to return back to the laboratory frame, giving\begin{equation}
U_{\Omega t}\tilde{H}_{\textrm{eff}}U_{\Omega t}^{\dagger}-i\hbar U_{\Omega t}\dot{U}_{\Omega t}^{\dagger}=\tilde{H}_{\textrm{eff}\left(2\right)}\,,
\label{H_eff-equivalence}
\end{equation} 
which is in agreement with Eq.~(\ref{eq:H_eff-xy}) for $H_{\textrm{eff}}$. Thus one arrives at completely equivalent effective Hamiltonians using both approaches.

\bibliography{Floquet-extra-space}

\end{document}